\documentclass[twocolumn,tighten]{aastex63}

\shorttitle{The MKID Exoplanet Camera for Subaru SCExAO}
\shortauthors{Walter et al.}

\graphicspath{{./}{figures/}}

\begin{document}

\title{The MKID Exoplanet Camera for Subaru SCExAO}

\correspondingauthor{Sarah Steiger}
\email{steiger@ucsb.edu}

\author{Alexander B. Walter}
\affiliation{Jet Propulsion Laboratory, California Institute of Technology, Pasadena, California 91125, USA}

\author{Neelay Fruitwala}
\affiliation{Department of Physics, University of California, Santa Barbara, CA 93111, USA}

\author[0000-0002-4787-3285]{Sarah Steiger}
\affiliation{Department of Physics, University of California, Santa Barbara, CA 93111, USA}

\author[0000-0002-4272-263X]{John I. Bailey, III}
\affiliation{Department of Physics, University of California, Santa Barbara, CA 93111, USA}

\author{Nicholas Zobrist}
\affiliation{Department of Physics, University of California, Santa Barbara, CA 93111, USA}

\author[0000-0001-5721-8973]{Noah Swimmer}
\affiliation{Department of Physics, University of California, Santa Barbara, CA 93111, USA}

\author[0000-0003-4792-6479]{Isabel Lipartito}
\affiliation{Department of Physics, University of California, Santa Barbara, CA 93111, USA}

\author[0000-0002-0849-5867]{Jennifer Pearl Smith}
\affiliation{Department of Physics, University of California, Santa Barbara, CA 93111, USA}

\author{Seth R. Meeker}
\affiliation{Jet Propulsion Laboratory, California Institute of Technology, Pasadena, California 91125, USA}

\author{Clint Bockstiegel}
\affiliation{Department of Physics, University of California, Santa Barbara, CA 93111, USA}

\author{Gregoire Coiffard}
\affiliation{Department of Physics, University of California, Santa Barbara, CA 93111, USA}

\author{Rupert Dodkins}
\affiliation{Department of Physics, University of California, Santa Barbara, CA 93111, USA}

\author{Paul Szypryt}
\affiliation{National Institute of Standards and Technology, Boulder, Colorado 80305, USA}
\affiliation{Department of Physics, University of Colorado, Boulder, Colorado 80309, USA}

\author[0000-0001-5587-845X]{Kristina K. Davis}
\affiliation{Department of Physics, University of California, Santa Barbara, CA 93111, USA}

\author{Miguel Daal}
\affiliation{Department of Physics, University of California, Santa Barbara, CA 93111, USA}

\author{Bruce Bumble}
\affiliation{Jet Propulsion Laboratory, California Institute of Technology, Pasadena, California 91125, USA}

\author{Giulia Collura}
\affiliation{Department of Physics, University of California, Santa Barbara, CA 93111, USA}

\author{Olivier Guyon}
\affiliation{National Astronomical Observatory of Japan, Subaru Telescope, Hilo, HI 96720}

\author{Julien Lozi}
\affiliation{National Astronomical Observatory of Japan, Subaru Telescope, Hilo, HI 96720}

\author{Sebastien Vievard}
\affiliation{National Astronomical Observatory of Japan, Subaru Telescope, Hilo, HI 96720}

\author[0000-0001-5213-6207]{Nemanja Jovanovic}
\affiliation{Caltech Optical Observatories, California Institute of Technology, Pasadena, CA 91125}

\author{Frantz Martinache}
\affiliation{Université Côte d'Azur, Observatoire de la Côte d'Azur, CNRS, Laboratoire Lagrange, France}

\author{Thayne Currie}
\affiliation{National Astronomical Observatory of Japan, Subaru Telescope, Hilo, HI 96720}
\affiliation{NASA-Ames Research Center, Moffett Blvd., Moffett Field, CA, USA}
\affiliation{Eureka Scientific, 2452 Delmer Street Suite 100, Oakland, CA, USA}

\author[0000-0003-0526-1114]{Benjamin A. Mazin}
\affiliation{Department of Physics, University of California, Santa Barbara, CA 93111, USA}

\begin{abstract}
We present the MKID Exoplanet Camera (MEC), a $z$ through J band (800--1400 nm) integral field spectrograph located behind The Subaru Coronagraphic Extreme Adaptive Optics (SCExAO) at the Subaru Telescope on Maunakea that utilizes Microwave Kinetic Inductance Detectors (MKIDs) as the enabling technology for high contrast imaging. MEC is the first permanently deployed near-infrared MKID instrument and is designed to operate both as an IFU, and as a focal plane wavefront sensor in a multi-kHz feedback loop with SCExAO. The read noise free, fast time domain information attainable by MKIDs allows for the direct probing of fast speckle fluctuations that currently limit the performance of most high contrast imaging systems on the ground and will help MEC achieve its ultimate goal of reaching contrasts of $10^{-7}$ at 2~$\lambda / D$. Here we outline the instrument details of MEC including the hardware, firmware, and data reduction and analysis pipeline. We then discuss MEC's current on-sky performance and end with future upgrades and plans. 
\end{abstract}

\keywords{MKIDs --- OIR Instruments --- Direct Imaging --- Adaptive Optics}

\section{Introduction} 
\label{sec:intro}

The discovery and characterization of exoplanets via direct imaging is particularly challenging due to the extreme contrasts ($<$10$^{-4}$ for ground based targets) and small angular separations ($\lesssim$1$^{\prime\prime}$) between the planetary companion and it's host star. For a review of the direct imaging of giant planets over the past few decades, see \citet{Bowler2016}.  Sophisticated ground-based high contrast instruments have been, or are being, built including: the Gemini Planet Imager \citep[GPI,][]{Macintosh2008, Macintosh2014}, SPHERE at VLT \citep{Carbillet2011, beuzit2019sphere}, SCExAO at Subaru \citep{Jovanovic2015}, P1640 \citep{Crepp2011, lewis2017direct} and the Stellar Double Coronagraph \citep[SDC,][]{Mawet2014} at Palomar, the Keck Planet Imager and Characterizer (KPIC) at Keck \citep{Mawet2016, jovanovic2019keck}, and MagAO-X at the Magellan Clay Telescope \citep{Males2018}. Adaptive optics (AO) and coronagraphy have enabled the discovery of planets with contrasts down to $\sim$10$^{-6}$. This has allowed researchers to survey formation conditions in primordial disks as well as the temperature of a handful of giant young exoplanets in the stellar neighborhood \citep{Marois+Macintosh+Barman+etal_2008, Lagrange+Bonnefoy+Chauvin+etal_2010, Kuzuhara+Tamura+Kudo+etal_2013, currie2013direct, Macintosh+Graham+Barman+etal_2015, Keppler+Benisty+Muller+etal_2018}. 

High-contrast imaging is limited primarily by uncontrolled diffracted light which produces a coherent speckle halo in the image plane. Speckles can arise from atmospheric aberrations, or from the non-ideal nature of the instrument, the latter suffering from noisy near-infrared (IR) detectors, time lag in the AO correction, speed of the AO control loop, non-common path errors, telescope vibrations, and chromaticity between the wavefront and science cameras \citep{Guyon_2005, Lozi_2018}. Atmospheric speckles are fast (rapidly evolving) and average down over an observation, while the slower, quasi-static speckles resulting from the non-ideal nature of the instrument must be removed using post-processing techniques. Angular differential imaging \citep[ADI,][]{Marois+Lafreniere+Doyon+etal_2006} exploits the rotation of the Earth or, analogously, the field-of-view of an altitude-azimuth telescope, to distinguish diffraction speckles, which remain stationary, from astrophysical sources, which will rotate with the frame. Spectral differential imaging \citep[SDI,][]{Racine+Walker+Nadeau+etal_1999,Marois+Doyon+Racine+etal_2000,Sparks+Ford_2002} uses the scaling of diffraction speckles with wavelength to distinguish speckles from true astrophysical sources. Since the initial development of ADI and SDI, a variety of post-processing algorithms have refined their approaches to push for higher achievable contrasts \citep[e.g.][]{Lafreniere+Marois+Doyon+etal_2007, Soummer+Pueyo+Larkin_2012, Marois+Correia+Galicher+etal_2014, Gonzalez_2016}. 

The time variability and chromaticity of quasi-static speckles, however, limit the performance of ADI and SDI which are the best performing current techniques \citep{Gerard+Marois+Currie+etal_2019}. Both of these techniques also suffer at small angular separations where exoplanets are more likely to occur. The speckle spectral dispersion used by SDI and the arclength traced by the companion’s sky rotation used by ADI, are both proportional to the planet’s separation. Furthermore, the precision of the background estimate for PSF subtraction is limited by low counting statistics at small separations \citep{Mawet_2014}. Even without these issues, the variability induced by speckle fluctuations can dominate over the shot noise expected from the total number of photons. 

A significant improvement over existing systems can be achieved by using read-noise free time domain information on short enough timescales to sample and control, or remove in post-processing, the fast speckle fluctuations from imperfections in the AO system and non-common path aberrations. We have constructed the MKID Exoplanet Camera (MEC) for Subaru Telescope's SCExAO system for this purpose - to serve as both a photon counting, energy resolving science camera, and as a focal plane wavefront sensor to correct for chromatic wavefront errors not accounted for by the upstream AO system. To achieve this, a high-speed, noise-free detector is required, like the Microwave Kinetic Inductance Detector (MKID) array used by MEC (\S\ref{sec:mkid}).

This instrument paper will highlight MEC's capabilities, the performance of the combined MEC and SCExAO system for high contrast imaging of exoplanets and disks, and document the significant new technology developed for MEC.

\section{MEC System Description}

MEC was commissioned at Subaru Observatory in 2018 and is designed as a $z$ through J-band IFU for high contrast imaging and as a focal plane wavefront sensor for a multi-kHz feedback loop with SCExAO. The enabling technology are Microwave Kinetic Inductance Detectors (MKIDs), which are capable of color discriminating near-IR photons with read noise free microsecond photon timing. The unique capabilities of MKIDs, coupled with SCExAO, provide a powerful means of overcoming the limits of ground based high contrast imaging and acts as a testbed for technologies aimed at future 30~m class telescopes. 

MEC's requirements were based on the delivered optical beam from SCExAO, the available space to mount MEC to SCExAO, the performance of the MKIDs, and the available budget. This resulted in a design with the instrument parameters that are summarized in Table \ref{tab:MEC_summary}. 

\begin{table}[!t]
  \centering
  \begin{tabular}{cc}
  \hline\hline
  Parameters & Values\\
  \hline
  Device Materials & PtSi on Sapphire w/ Nb g.p.\\
  Device Format & 140x146 pixels (10 feedlines)\\
  Pixel Pitch & 150 $\mu$m\\
  Plate Scale & 10.4 mas/pixel\\
  Field of View & 1.4" x 1.5"\\
  Wavelength Band & 800-1400~nm ($z$ - J)\\
  Spectral Resolution ($\lambda / \Delta \lambda$) & 5-7\\
  Max Count Rate & 5000 cts/pix/second\\
  Pixel Dead Time & 10~$\mu$s\\
  Readout Frame Rate & $>$2~kHz\\
  Operating Temp. & 90~mK\\
  4~K Stage Base Temp. & 3.1~K\\
  60~K Stage Base Temp. & 57~K\\
  90 mK Hold Time & $>$17 hours \\
  \hline
  \end{tabular}
  \caption{MEC Instrument Summary}  \label{tab:MEC_summary}
  \vspace{-12pt}
\end{table}

\subsection{Optical and Infrared (OIR) MKID Arrays}
\label{sec:mkid}

MKIDs work on the principle that incident photons change the surface impedance of a superconductor through the kinetic inductance effect. The kinetic inductance effect occurs because energy can be stored in the supercurrent (the flow of Cooper Pairs) of a superconductor. Reversing the direction of the supercurrent requires extracting the kinetic energy stored in it, which yields an extra inductance term in addition to the familiar geometric inductance. The magnitude of the change in surface impedance depends on the number of Cooper Pairs broken by incident photons, and hence is proportional to the amount of energy deposited in the superconductor. This change can be accurately measured by placing a superconducting inductor in a lithographed resonator. A microwave probe signal is tuned to the resonant frequency of the resonator and any photons which are absorbed in the inductor will imprint their signature as changes in the phase and amplitude of this probe signal. 

\begin{figure}[!t] 
  \includegraphics[width=\columnwidth]{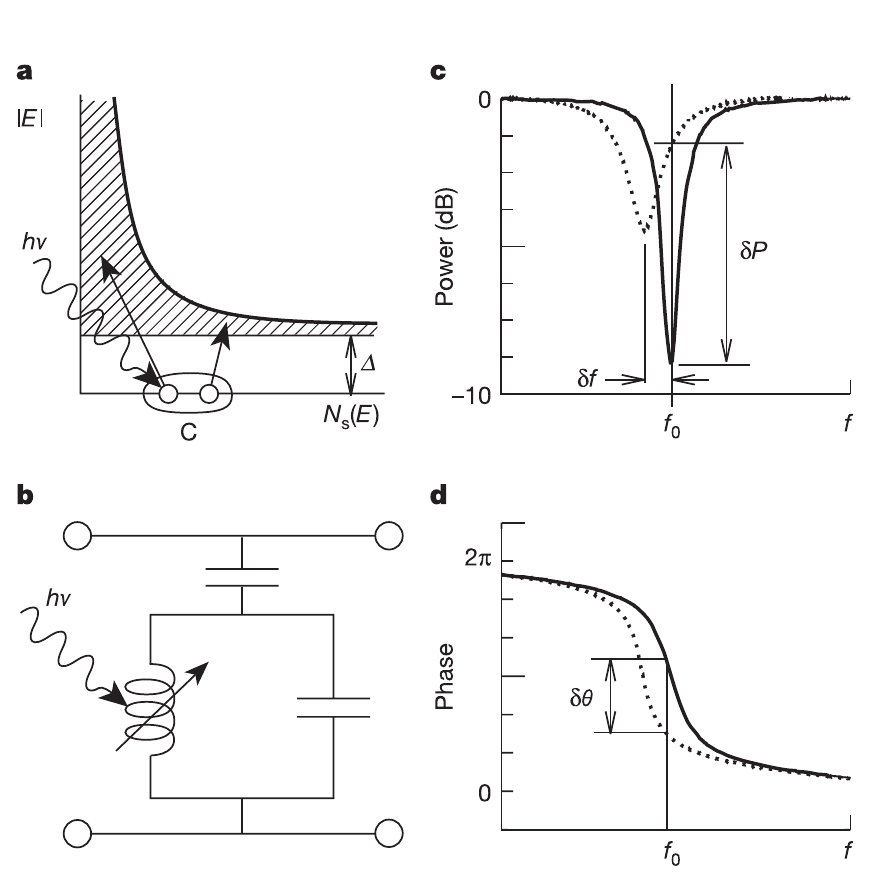}
  \caption{The basic operation of an MKID, from \citet{Day_2003}. (a) Photons with energy $h\nu$ are absorbed in a superconducting film producing a number of excitations called quasiparticles. (b) To sensitively measure these quasiparticles the film is placed in a high frequency planar resonant circuit. The amplitude (c) and phase (d) of a microwave excitation signal sent through the resonator. The change in the surface impedance of the film following a photon absorption event pushes the resonance to lower frequency and changes its amplitude. If the detector (resonator) is excited with a constant on-resonance microwave signal, the energy of the absorbed photon can be determined by measuring the degree of phase and amplitude shift.}
  \label{fig:mkid_operating}
\end{figure}

Since the quality factor, Q, of the resonators is high, and their transmission off resonance is nearly perfect, multiplexing can be accomplished by tuning each pixel to a different resonant frequency with lithography during device fabrication. A comb of probe signals is sent into the device and room temperature electronics recover the changes in amplitude and phase. This makes a device capable of measuring the arrival time (to a microsecond) and energy (to 5-10\%) of each arriving photon without read noise or dark current; an optical/near-IR analog of an X-ray microcalorimeter. More details on MKIDs can be found in \citet{Mazin_2012, ARCONS} and in Figure \ref{fig:mkid_operating}.

\begin{figure}[!t] 
  \includegraphics[width=\columnwidth]{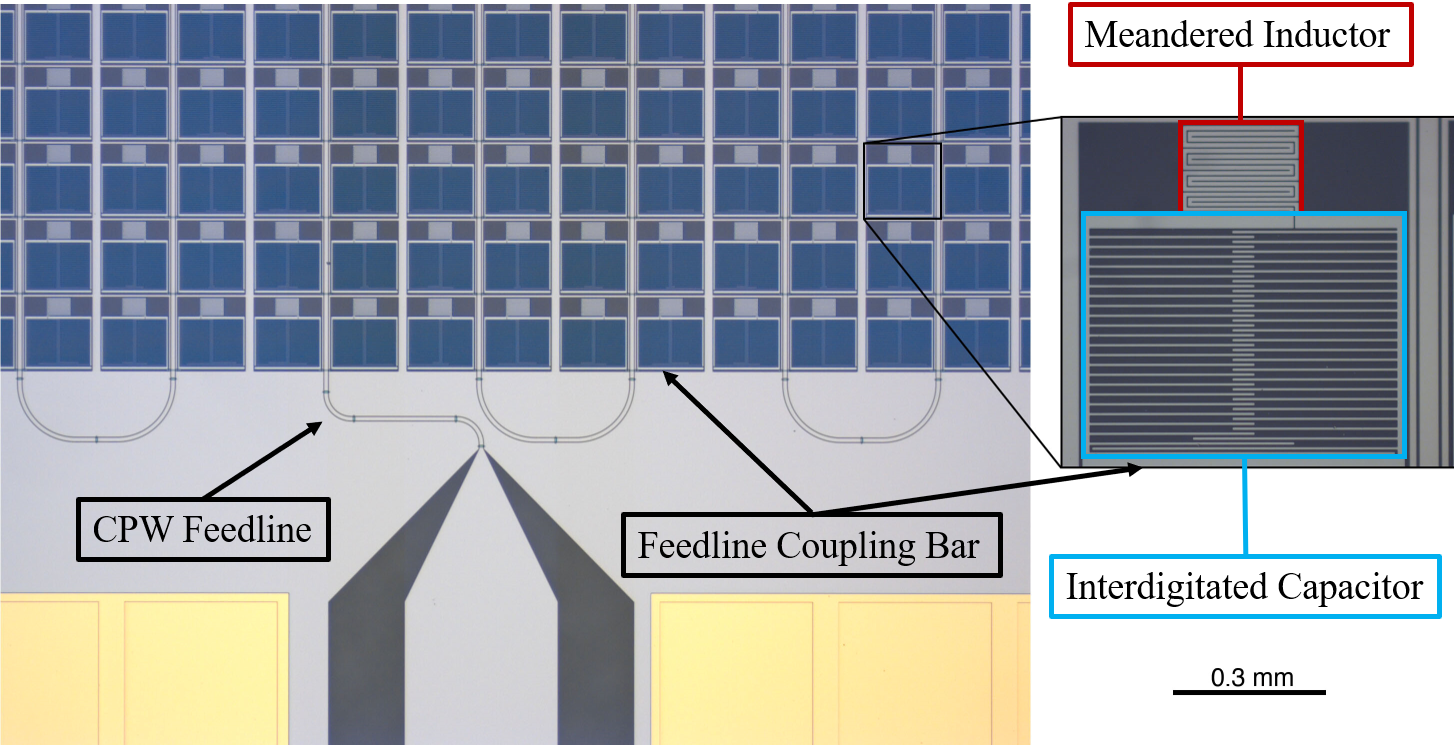}
  \caption{Microscope image of a small section of a 20 kpix MKID array similar to the engineering array in the MEC instrument. Left: A coplanar waveguide (CPW) feedline with crossovers connects to a grid of MKID pixels. Gold thermal pads are shown at the bottom. Right: Zoomed in picture showing the large interdigitated capacitor, meandered inductor, and the feedline coupling bar (bottom right corner).}
  \label{fig:mkid_image}
  \vspace{-2pt}
\end{figure}

The MEC MKID array has 10 coplanar waveguide (CPW)  transmission lines, or feedlines, that each probe 14 x 146 pixels for a total of 20440 pixels in the array. This makes it the largest superconducting detector in the world. It uses the PtSi MKIDs described in \citet{Szypryt_2017}, which are optimized for 800-1400~nm light with a spectral resolution of about $\lambda / \Delta \lambda=6$. The pixels are on a 150~$\mu m$ pitch, see Figure \ref{fig:mkid_image}. The device currently in MEC has 7/10 feed lines with good transmission and on those there is about an 80\% pixel yield.  More detail on current performance can be found in \S\ref{sec:perf}. 

\begin{figure}[!ht] 
  \vspace{0.2in}
  \centering
  \includegraphics[width=\columnwidth]{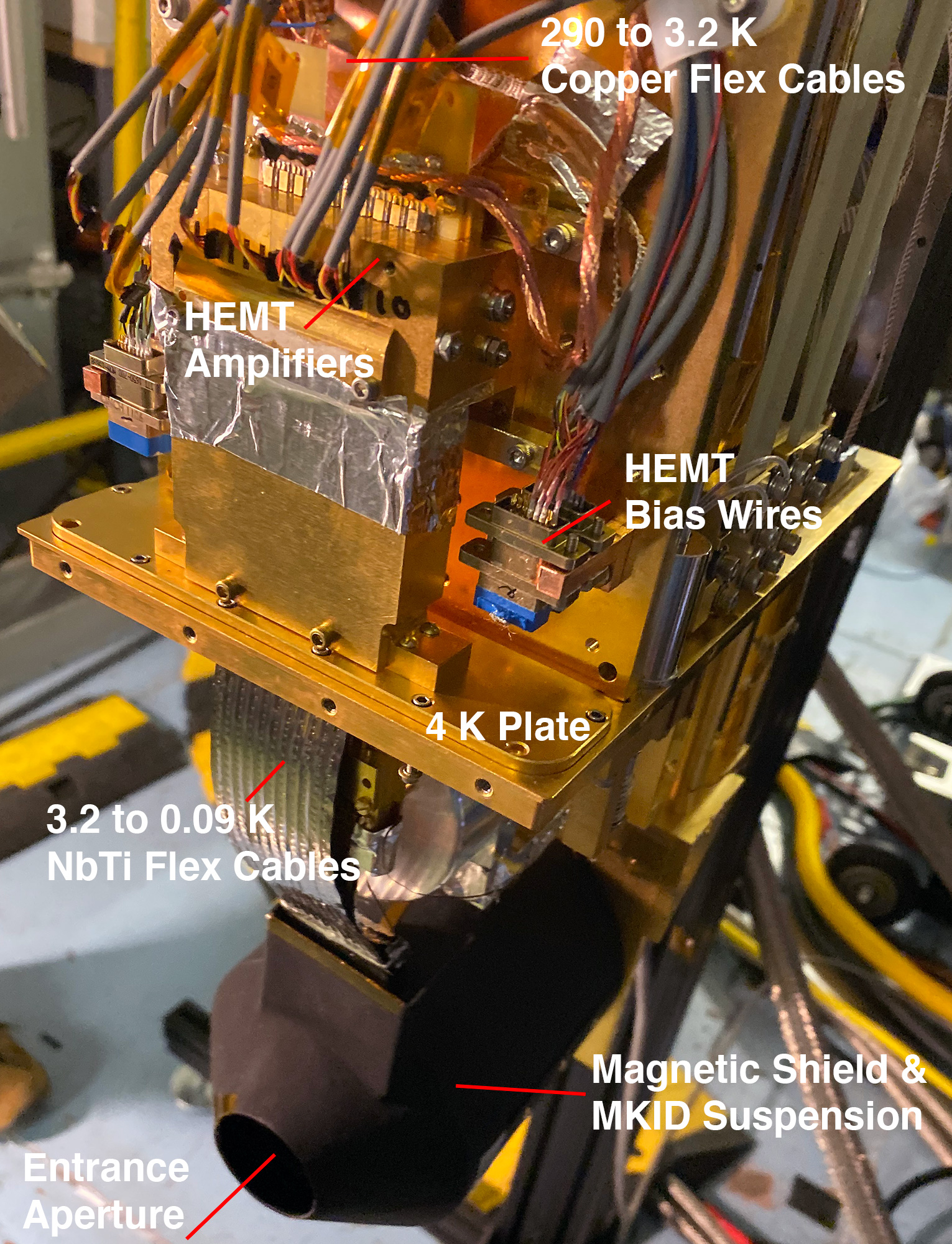}
  \caption{Image of the 4 K stage of MEC with the radiation shields removed to show the readout wiring and the black magnetic shield. The pulse tube and heat switch are hidden by the wiring attachment plate. }
  \label{fig:MECcold}
\end{figure}

\begin{figure}[!ht] 
  \vspace{0.2in}
  \centering
  \includegraphics[width=\columnwidth]{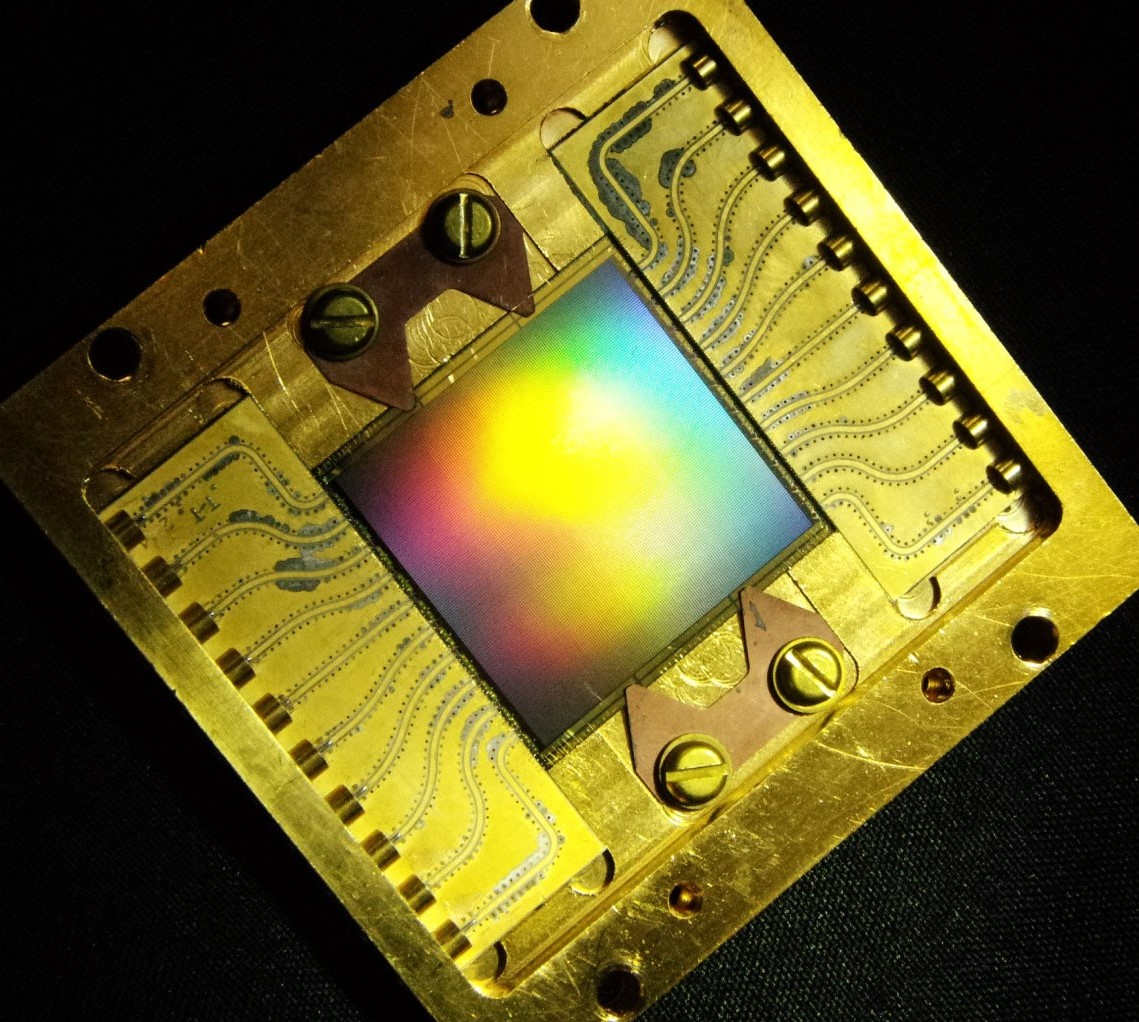}
  \caption{Image of a 20440 pixel PtSi MKID device designed for MEC. This is the highest pixel count superconducting detector array at any wavelength.}
  \label{fig:mkid_device_image}
\end{figure}

\subsection{Cryostat}

MEC's cryostat is a pulse tube cooled Adiabatic Demagnetization Refrigerator (ADR) capable of reaching temperatures below 50~mK. The outer vacuum shell 300~K enclosure measures roughly 22~cm wide x 33~cm deep x 96~cm tall. It contains RF gaskets to reduce radio frequency interference, a 7.5~kW pulse tube, and all the vacuum and wiring connections to accommodate a 20 kpix MKID array. 


The 60~K shell is cooled by the first stage of the pulse tube. It contains heat sinks to cool the wiring to reduce heat load on the 4~K stage. The 4~K stage is cooled by the second stage of the pulse tube, and has 0.75~W of cooling power at 4~K. 

Attached to the 4~K plate, shown in Figure~\ref{fig:MECcold} are: the ADR unit with mechanical heat-switch, the ten cryogenic high-electron-mobility transistor (HEMT) amplifiers, and the detector package. The detector package is mounted to the 4~K plate and is enclosed in a 4~K magnetic shield. This design places the MKID array far from the magnetic shield opening where field leakage will be strongest and also allows us to install a 1~K, 9~cm long black baffle to further reduce off-axis scattered light and 4~K black body radiation. The MKID array is mounted in a microwave package 5~cm x 5~cm x 0.5~cm in size, as shown in Figure \ref{fig:mkid_device_image}. This box is attached to a gold-plated copper rod that sticks out of the base of the magnetic shield and connects to the ADR unit by a copper strap. 

The ADR acts as a single-shot magnetic cooler which brings the MKID array down to 90 mK where the temperature is stabilized with a feedback loop to the ADR magnet power supply. We achieve a 90 mK hold time of $>$17 hours on the telescope, more than sufficient for the 13-14 hours required for a night of calibration and on-sky observations.

\subsection{Readout}
\subsubsection{Readout Electronics}
\label{sec:readout_electronics}

\begin{figure*}[!ht] 
  \begin{center}
  \includegraphics[width=0.5\textwidth]{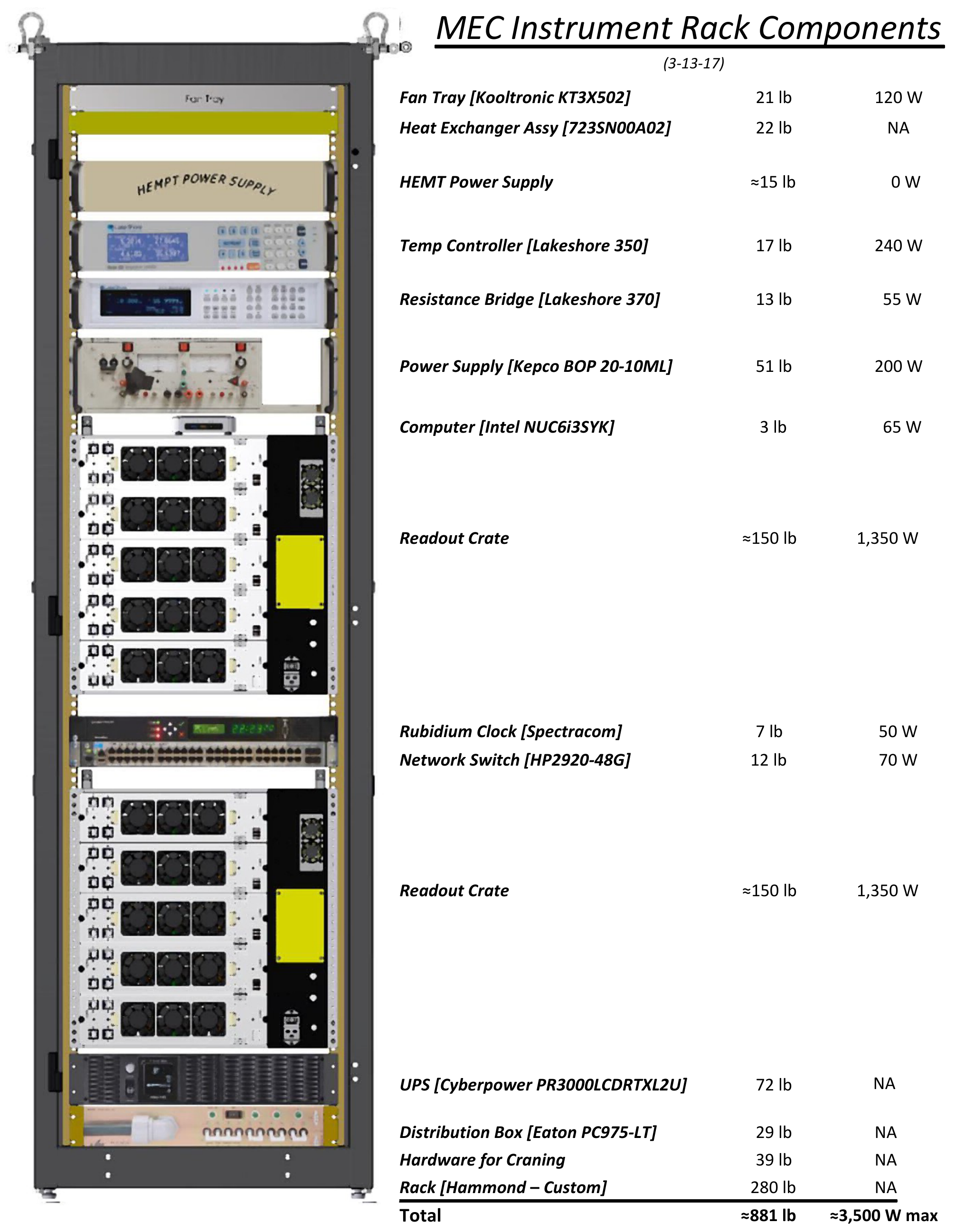}
  \end{center}
  \caption{CAD drawing of MEC electronics rack (E-rack) and list of components. }
  \label{fig:mec_erack}
\end{figure*}
\vspace{-4pt}

\begin{figure}[!t]  
  \begin{center}
  \includegraphics[width=\columnwidth]{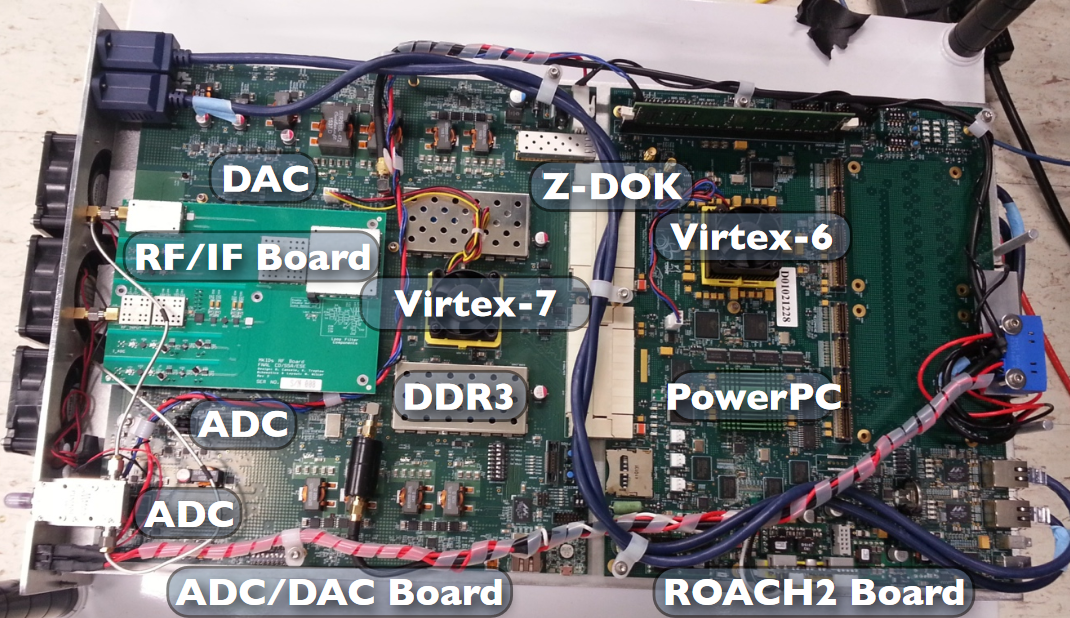}
  \end{center}
  \caption{The ROACH2 board is connected to the ADC/DAC board by two Z-DOK connectors. The RF/IF board is mounted on the ADC/DAC board using SMP blind-mate connectors for signals and general-purpose input/output (GPIO) pins for programming. Another set of three boards are mounted to the underside of this cartridge. Figure and Caption reproduced from \citet{Strader_2016}.}
  \label{fig:readout}
\end{figure}

The electronics rack (E-rack) is situated on the Nasmyth platform next to the MEC instrument (Figure \ref{fig:mec_erack}). It requires 3.5 kW max of power supplied to the power distribution unit with a 30~A, 208~V 3-phase power cord. Cooling water/glycol is supplied to the E-rack by Subaru. There should be unobstructed air flow to two readout crates in the electronics rack. The front (facing SCExAO) draws air in, and the back expels air through a heat exchanger into the Nasmyth room.

\begin{figure*}[!t] 
  \begin{center}
  \includegraphics[width=\textwidth]{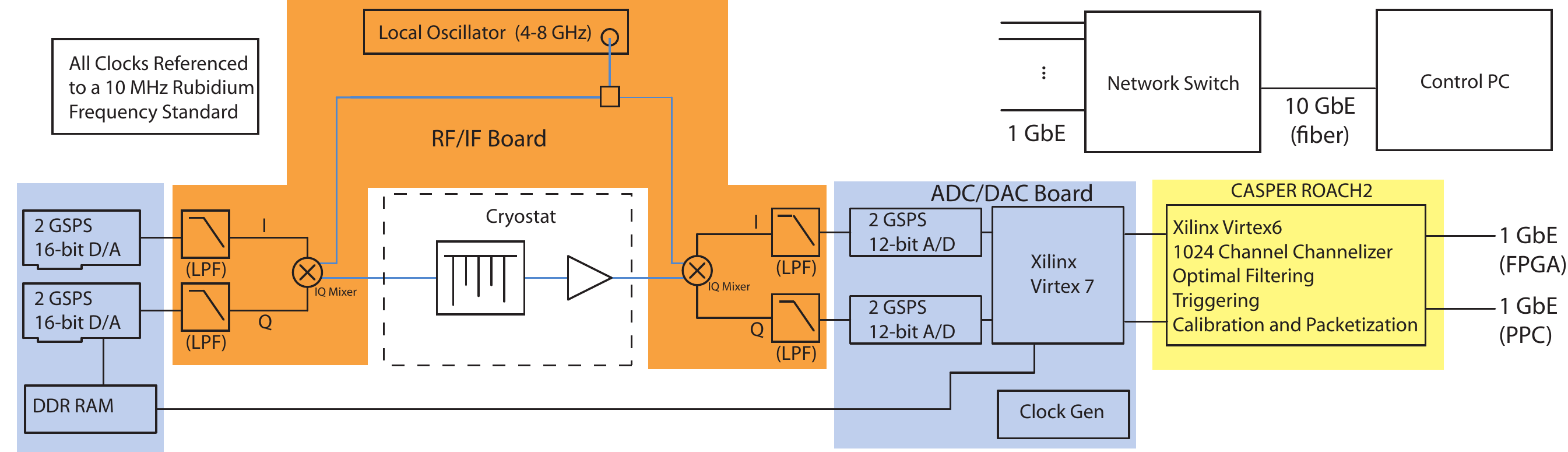}
  \end{center}
  \caption{Readout system block diagram. This diagram is for 1 set of boards which can read out 1024 pixels at a time. Figure reproduced from \citet{Meeker2018}}
  \label{fig:firmware}
\end{figure*}

The readout for MEC is based on the ARCONS readout, which is detailed in \citet{McHugh_2012} and is identical to the DARKNESS readout detailed in \citet{Strader_2016, NeelayGen2}. The readout system is responsible for generating the RF frequency comb to drive the resonators, and for digitizing and processing the resulting MKID array output to determine the arrival time  energy of any incident photons,  see \S\ref{sec:readout_procedure}. The MEC readout employs twenty second generation CASPER Reconfigurable Open Architecture Computing Hardware (ROACH2) boards. Each of these ROACH2 boards are connected to an Analog to Digital/Digital to Analog Converter (ADC/DAC) board and a Radio Frequency/Intermediate Frequency (RF/IF) board, both of which are designed at Fermilab (See Figure \ref{fig:readout} for one board set). Each set of boards reads out 1024 pixels in 2 GHz of bandwidth.

Every ROACH2 houses a Xilinx Virtex6 Field Programmable Gate Array (FPGA), which processes and channelizes the signal from the ADC using a scaled up version of the ARCONS firmware. The ADC/DAC board houses a Virtex7 FPGA. The ROACH2 and ADC/DAC boards are connected by two Z-Dok connectors, which are responsible for sending the ADC ouptut and clock signals to the ROACH2, and for implementing an SPI communications interface between the Virtex-6 and Virtex-7 FPGAs.

The ROACH2 boards are on a private local-area network (LAN) as moderated by the network switch. We run all readout control software on the MEC data server, however, any computer (with the required Python libraries) that is connected to the LAN can perform this task. We use a PPS (pulse per second) signal generator to sync the clocks for each set of readout boards. The PPS signal is sychronized to an external PTP/NTP server to ensure accurate absolute timing as well as synchronization with the data server clock (See Figure \ref{fig:firmware}).

\subsubsection{Readout Procedure}
\label{sec:readout_procedure}

The initial setup defines the frequency comb that probes the resonators. This is determined by doing frequency sweeps of every feedline at a grid of powers and then using a machine learning code to both find the resonators and determine their optimal probe powers. An algorithm to find the probe power of known resonators is outlined in \citet{2018A&C....23...60D}, but has been superseded by an algorithm that simultaneously finds resonances and their ideal power which will be detailed in an upcoming publication. Our python software then computes a lookup table (LUT) containing the time-domain sum of the resonator probe tones. The lookup table is stored in DDR3 RAM onboard the ADC/DAC board, and is played to the DAC in a loop while the system runs. The DAC output is passed through the RF/IF board where an IQ mixer upconverts the signal to the right frequency range (4 to 6 GHz or 6 to 8 GHz) using a local oscillator. After traveling through the cryostat, MKID array, and HEMTs, the RF/IF board mixes down the returning signal to baseband where it is digitized by the ADCs. Then the Virtex7 on the ADC/DAC board streams the raw ADC output to the ROACH2 over the ZDOK connectors. The ROACH2 firmware processes ADC data stream, performing the following actions:
\begin{enumerate}
    \item Separate the comb of frequencies into individual pixel frequencies using an FFT (Fast Fourier Transform) and digital down conversion (DDC).
    \item Sample the phase of each resonator's signal every microsecond.
    \item Run a pixel-wise unique optimal filter over each pixel's phase signal.
    \item Trigger on photon events and store photon packets in buffer.
    \item Send buffer to disk every 0.5 millisecond (or faster if full) over 1-Gbit Ethernet via UDP.
\end{enumerate}

Each single photon packet is a 64 bit word with the following breakdown: 10 bit x-coordinate, 10 bit y-coordinate, 9 bit timestamp, 18 bit wavelength, 17 bit baseline phase. A C program on the data server receives the UDP frames containing photon packets sent from the ROACH2 boards and writes them to disk as observation files. Additionally, it histograms the photons into images for real time display. 

\subsection{Optics}

\subsubsection{Specifications of Incoming Beam}
The SCExAO exit beam is located 293$\pm$1~mm from the top of the top mounting rail. The beam comes off the bench at about 58~mm from the left edge of the bench. The SCExAO exit beam is collimated. The pupil position is 54$\pm$20~mm from the back of the bench. The pupil size is 8.67$\pm$0.15~mm. The beam is expected to exhibit $<$10~nm of chromatic RMS optical wavefront error over the 0.8-1.4 $\mu$m bandwidth.

\subsubsection{MEC Fore-Optics}
\begin{figure}[!t] 
  \includegraphics[width=\columnwidth]{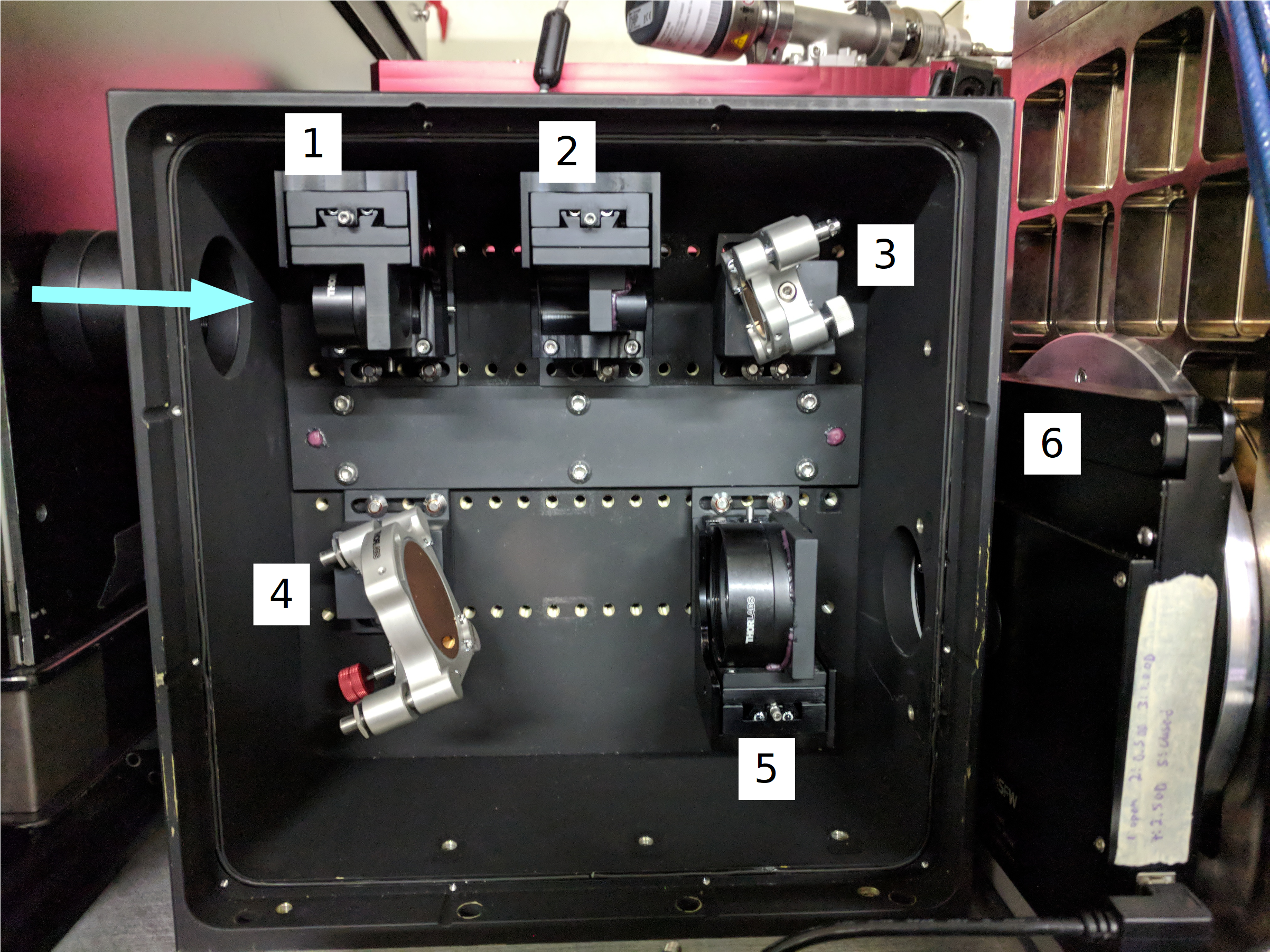}
  \caption{The MEC foreoptics box reimages the beam from SCExAO (arrow from left) onto the MKID device. Two gold mirrors (3 and 4) and three lenses (1, 2 and 5) steer and focus the input collimated beam onto the MKID array with an f-number of f/377.7. The specific optics components can be found with their corresponding numbers in Table \ref{tab:MEC_optics}.}
  \label{fig:mec_foreoptics}
\end{figure}

The MEC optics box has a two inch hole for the incoming SCExAO beam (Figure \ref{fig:mec_foreoptics}). A set of three lenses creates a telecentric beam designed to reimage the focus onto the MKID array as simulated in Zemax. The beam leaves the optics box with an f-number $f/377.7$, travels through the neutral density (ND) filter wheel (Table \ref{tab:ND_filters}), and enters the cryostat through the front vacuum window. There are two IR bandpass filters at the 60~K and 4~K shield to block black body radiation and finally a microlens array focuses the beam onto the light sensitive part of the MKID pixels. The MEC optics are listed in Table \ref{tab:MEC_optics}. 

\begin{table}[!t]
  \centering
  \caption{MEC Optics}
  \label{tab:MEC_optics}
  \begin{tabular}{cll}
  \hline\hline
  \# & Optic & Description\\
  \hline
  1 & 100~mm FL lens & Thorlabs AC254-100-c\\
  2 & 9~mm FL lens & Edmunds 45-783\\
  3 & 2.54~mm gold mirror & Mnt. on CONEX-AG-M100D\\
  4 & 5.08~mm gold mirror & Fixed mount\\
  5 & 300~mm FL lens & Thorlabs AC508-300-c\\
  6 & 2~in ND Filter wheel & See Table \ref{tab:ND_filters}\\
  7 & Front Window & Edmund 48-130\\
  8 & 60~K IR filter & 10~mm thick N-BK7\\
  9 & 4~K IR filter & 20~mm thick N-BK7\\
  10 & Microlens Array & a$\mu$s APO-GT-P150-R0.8\\
  \hline

  \end{tabular}
\end{table}
\begin{table}[!t]
  \centering
  \caption{MEC Neutral Density Filter Wheel}
  \label{tab:ND_filters}
  \begin{tabular}{ccl}
  \hline\hline
  Slot & Filter & Part \#\\
  \hline
  1 & OPEN &\\
  2 & 0.5 OD & Edmund 62-673\\
  3 & 1.0 OD & Edmund 62-676\\
  4 & 2.5 OD & Edmund 62-680\\
  5 & CLOSED &\\
  \hline

  \end{tabular}
\end{table}

\subsection{Integration with SCExAO}

\begin{figure}[!t] 
  \centering
  \includegraphics[width=\columnwidth]{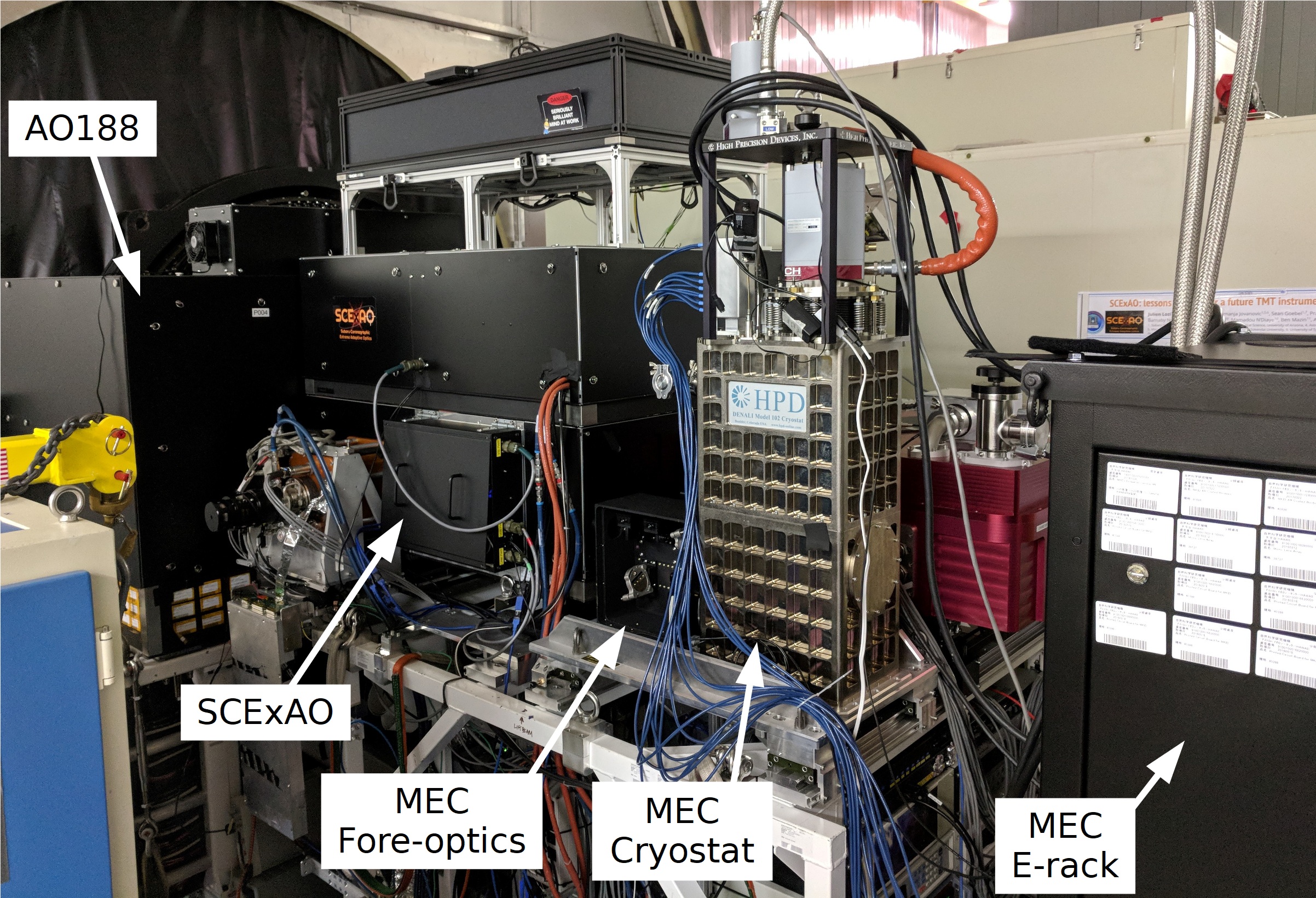}
  \caption{MEC is mounted behind the SCExAO bench and AO188 facility adaptive optics systems on the IR Nasmyth port at Subaru Telescope. }
  \label{fig:mec_nasmyth_image}
\end{figure}

\begin{figure}[!t] 
  \centering
  \includegraphics[width=\columnwidth]{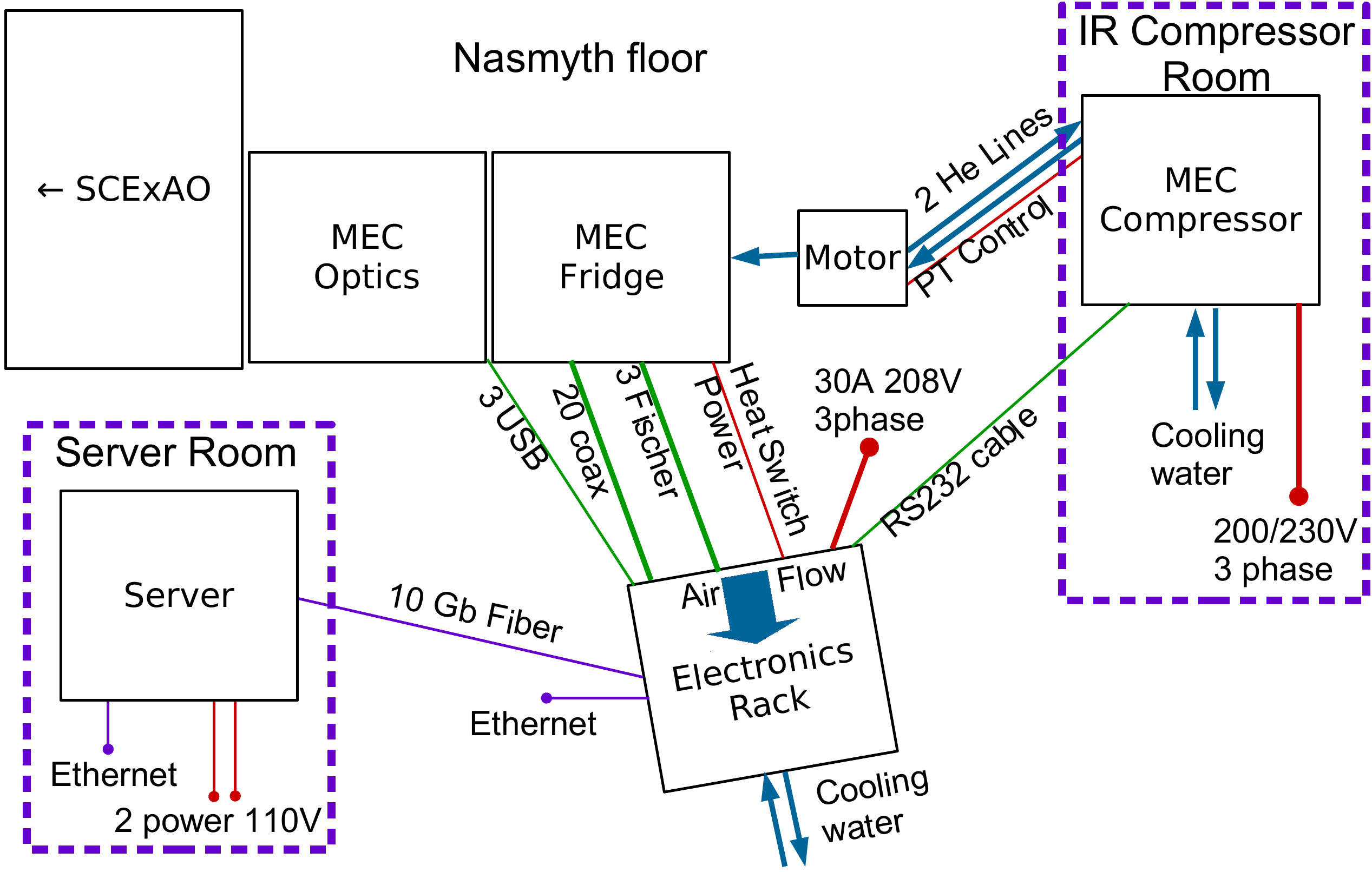}
  \caption{A diagram showing the MEC floor plan and interconnects at the Subaru Telescope. Here, red lines denote power connections, blue lines/arrows denote cooling lines, green lines denote electrical cabling and purple lines denote network connections. }
  \label{fig:mec_floorplan}
\end{figure}

\begin{figure*}[!t] 
  \begin{center}
  \includegraphics[width=\textwidth]{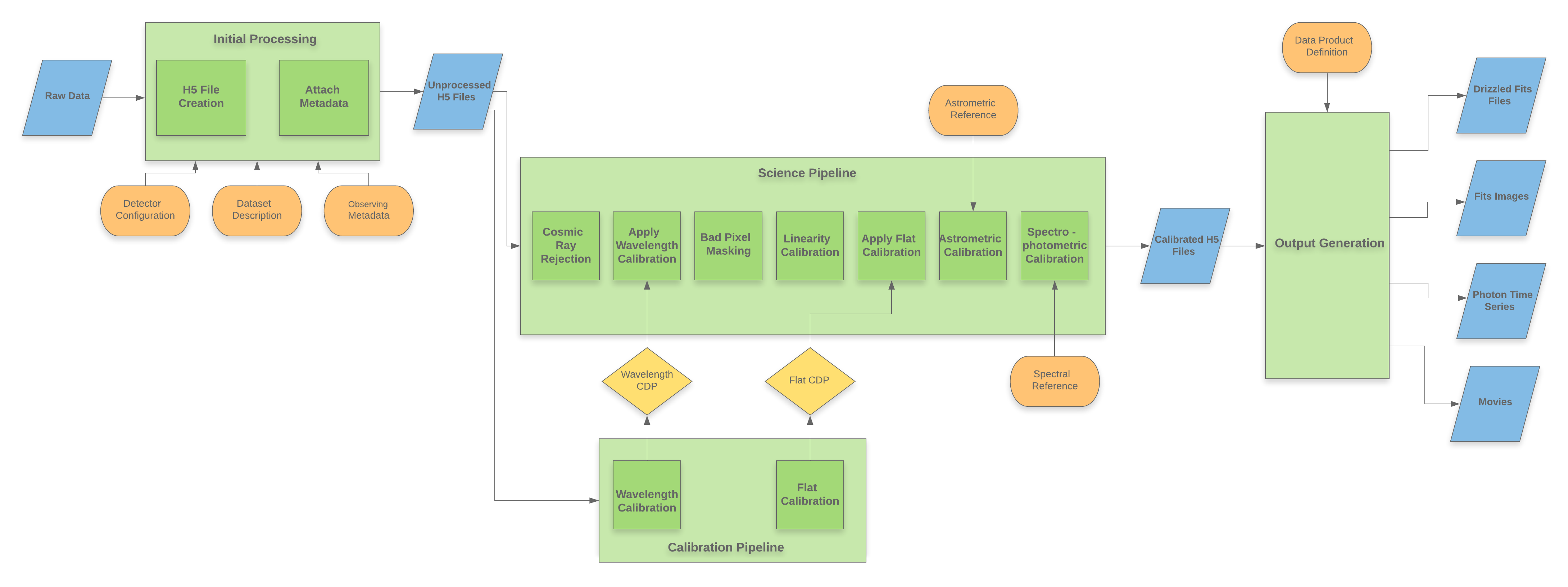}
  \end{center}
  \caption{Block Diagram of The MKID Pipeline. Calibration steps are shown in green, references and configurations are shown in orange, data products are shown in blue, and calibration data products (CDP) are shown in yellow.}
  \label{fig:pipeline}
\end{figure*}

As of March 19, 2018, MEC is mounted on the Nasmyth platform behind SCExAO in the place of the High-Contrast Coronographic Imager for Adaptive Optics (HiCIAO; \cite{hodapp2008hiciao}). The E-rack is stationed next to MEC on the Nasmyth floor. A He compressor is located in the IR compressor room on the observation floor and a linux server is in the server room on the first floor of the control building. Figure \ref{fig:mec_floorplan} diagrams how MEC is situated on the Nasmyth floor. Figure \ref{fig:mec_nasmyth_image} is an image of MEC mounted at the telescope. 
\subsection{Data Processing/Calibration}

To analyze MEC data we use The MKID Pipeline available on Github: \url{https://github.com/MazinLab/MKIDPipeline}.

The MKID Pipeline begins by transforming the raw binary data into calibrated photon lists in a fast access database. This process is outlined in Figure \ref{fig:pipeline}. Many of the steps were implemented for the ARCONS pipeline \citep{Eyken_2015} and were copied or slightly modified for MEC. 

\begin{enumerate}
    \item Processing Binary Data into a HDF5 Database
    \begin{itemize}
        \item[] The raw binary observation files, consisting of header and photon data sent over ethernet, are interpreted and saved into a HDF5 table format. The table has one entry for each photon with the following fields: pixel ID, arrival time, phase height (or wavelength), and two additional weight fields to be set in later calibration steps (spectral weight and noise weight). The photon list is rearranged so that it is first ordered by pixel location and then by time. 
    \end{itemize}
    \item Cosmic Ray Rejection
    \begin{itemize}
        \item[] Cosmic rays incident on an MKID detector have the effect of illuminating most of the array with false high energy photon counts. This calibration step removes time chunks from further analysis that are identified as containing a cosmic ray. The advantage of MKID's timing resolution is that these time chunks can be precisely selected to remove an average of only 0.01\% of a given observation. This retains as much of the original exposure time as possible. 
    \end{itemize}
    \item Wavelength Calibration
    \begin{itemize}
        \item[] The wavelength calibration is essential to utilize MKID's energy resolving capabilities. Several monochromatic laser exposures are used to determine a quadratic, $a^2\phi + b\phi +c = hc/\lambda$, that converts phase, $\phi$, into wavelength, $\lambda$ for each pixel. To save space, the phase height float column is overwritten with the calibrated wavelength for each photon that has a valid wavelength solution. 
    \end{itemize}
        \item Bad Pixel Masking
    \begin{itemize}
        \item[] Hot, cold, or otherwise badly behaving pixels are prevalent in our data since we try to include as many pixels as possible from the readout step. These pixels are identified by determining whether their measured flux value deviates from the median flux of pixels in a surrounding box by a user specified number of standard deviations. These pixels can then be removed from the final data product. 
    \end{itemize}
        \item Linearity Correction
    \begin{itemize}
        \item[] This calibration accounts for missing photons due to the detector dead time. This dead time (10 $\mu s$ for MEC) is introduced in the firmware to avoid photon pile-up and limit the total count rate, but results in a non-linear detector response at high count rates. To account for this effect, a correction factor is applied to the spectral weight column for each photon depending on the local count rate.
    \end{itemize}
    \item Flat Field Calibration
    \begin{itemize}
        \item[] This is a standard flat field calibration that determines the inter-pixel relative QE. This calibration is done as a function of wavelength and can be achieved using either a white light source, or the monochromatic calibration data used for the wavelength calibration. The resulting flat weight multiplies into the spectral weight column of the HDF5 table to merge with the linearity weight for each photon.
    \end{itemize}
      \item Astrometric Calibration
    \begin{itemize}
        \item[] This is an implementation of a standard astrometric calibration to relate (x,y) pixel coordinates on the MKID array to sky coordinates using a linear mapping with rotation.  
    \end{itemize}
    \item Spectral Calibration
    \begin{itemize}
        \item[] This calibration accounts for the total QE of the detector and system as a function of wavelength. A spectrum is taken of a spectral standard with MEC and is compared to a calibrated spectrum of the same object. The resulting correction multiplies into the spectral weight column of the HDF5 table to merge with the flat weight and linearity weight for each photon.
    \end{itemize}
    \item Output Generation
    \begin{itemize}
        \item[] There are a variety of output file formats that the MKID Pipeline can generate. In addition to FITS files which are compatible with more standard post-processing techniques, photon lists can be output directly to perform more specialized techniques that utilize the photon-counting ability of MKIDs.  
    \end{itemize}
\end{enumerate}



\subsubsection{Analysis of Calibrated Data}
After the photon lists have been calibrated we can begin the analysis of astrophysical data produced by MEC. The easiest to understand procedure is to bin the photon list in time and wavelength to create images after which classical astronomy post-processing techniques can be used. 

\begin{figure}[!t] 
  \centering
  \includegraphics[width=\columnwidth]{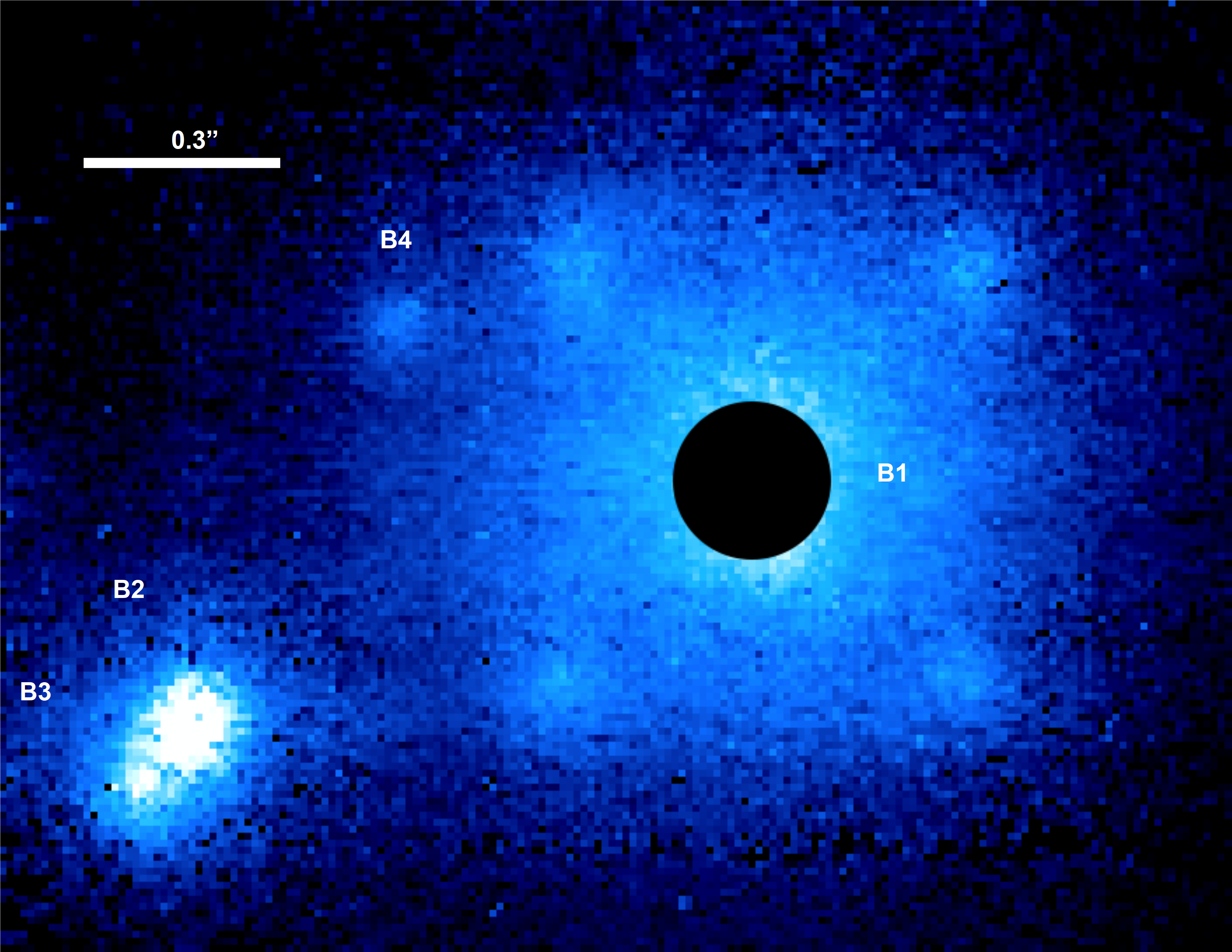}
  \caption{y-band image of Theta$^1$ Orionis B taken by MEC at the Subaru Telescope on Jan. 12, 2019. A dither sequence of 25 30~s images are stacked and then smoothed by a Gaussian filter. The black circle indicates the location of the coronagraph on B1. The four astrogrid speckles artificially created by the DM are present in a square centered about, and about 5.5 magnitudes fainter than, B1. B4 is 4.98 magnitudes fainter in H band than B1. The four artificial speckles are for astrometric and photometric calibration. }
  \label{fig:trap}
\end{figure}

Figure \ref{fig:trap} shows the five star system Theta$^1$ Orionis B taken in y-band by MEC at Subaru Telescope on Jan. 12, 2019. B1, behind the coronagraph, is a spectroscopic eclipsing binary with nominal magnitude of V=7.96. B2 and B3 are resolved by Subaru Telescope's 8.2~m mirror in conjunction with the AO correction provided by SCExAO \citep{Jovanovic2015}. The four astrogrid speckles artificially created by the DM are present in a square centered about B1, behind the coronagraph. These speckles are about 5.5 magnitudes fainter than B1. B4 is 4.98 magnitudes fainter than B1 in H-band \citep{Close_2013}. It is important to note that this observation was taken before the re-coating of the secondary mirror at Subaru which occurred on November 7th, 2019. The throughput hit taken before the mirror re-coating, coupled with the already dim guide star, B1, leads to a worse AO correction than is possible either now on the same system, or on a similar system with a brighter guide star. 

The image uses data from 25 different 30~s observations located at different positions on-sky collected in a dither. We use a drizzling code to place the list of observed photons in a virtual pixel grid.  that correctly takes into account the translation from the dither position as well as sky rotation as observed from the Nasmyth port. This helps to account for the lack of full array coverage due to dead feedlines and pixels to generate a complete image.

It is also possible to use the data directly without any binning in the calibrated photon list for more complex analysis, like the stochastic speckle discrimination technique outlined in~\citet{Walter2019a}.

\section{MEC On-Sky Performance}
\label{sec:perf}

MEC underwent a major upgrade in late 2019 where we replaced internal microwave wiring that was degrading resonator performance due to crosstalk and excess noise~\citep{2020arXiv200706496S}.  The results below reflect the performance after the upgrade. Due to fabrication problems we have also not yet been able to improve on the engineering grade MKID array in MEC, although we hope to install a science grade MKID array with all feedlines functional and higher yield in 2020.  Luckily, most of the exoplanets of interest fit within the smaller field of view of the engineering grade array, although the full array will feature a 1.4" x 1.5" field of view to improve observations of disks.

\subsection{Yield}

Yield, or the number of functioning pixels compared to the total number of possible pixels, can be complex in an MKID array as every pixel is unique and can be compromised in a number of ways. For example, a spec of dust landing on a pixel during fabrication can short out a resonator, or merely move it to an unpredictable frequency. Non-uniformities in film thickness or composition can move resonators around in frequency, causing frequency overlaps that can render at least one of the pixels unable to be read out. Even worse, fabrication problems such as shorts due to photoresist bubbles can knock out entire feedlines, as seen in feedlines 2--4 in the MEC engineering array.

\begin{figure}[!t] 
  \centering
  \includegraphics[width=\columnwidth]{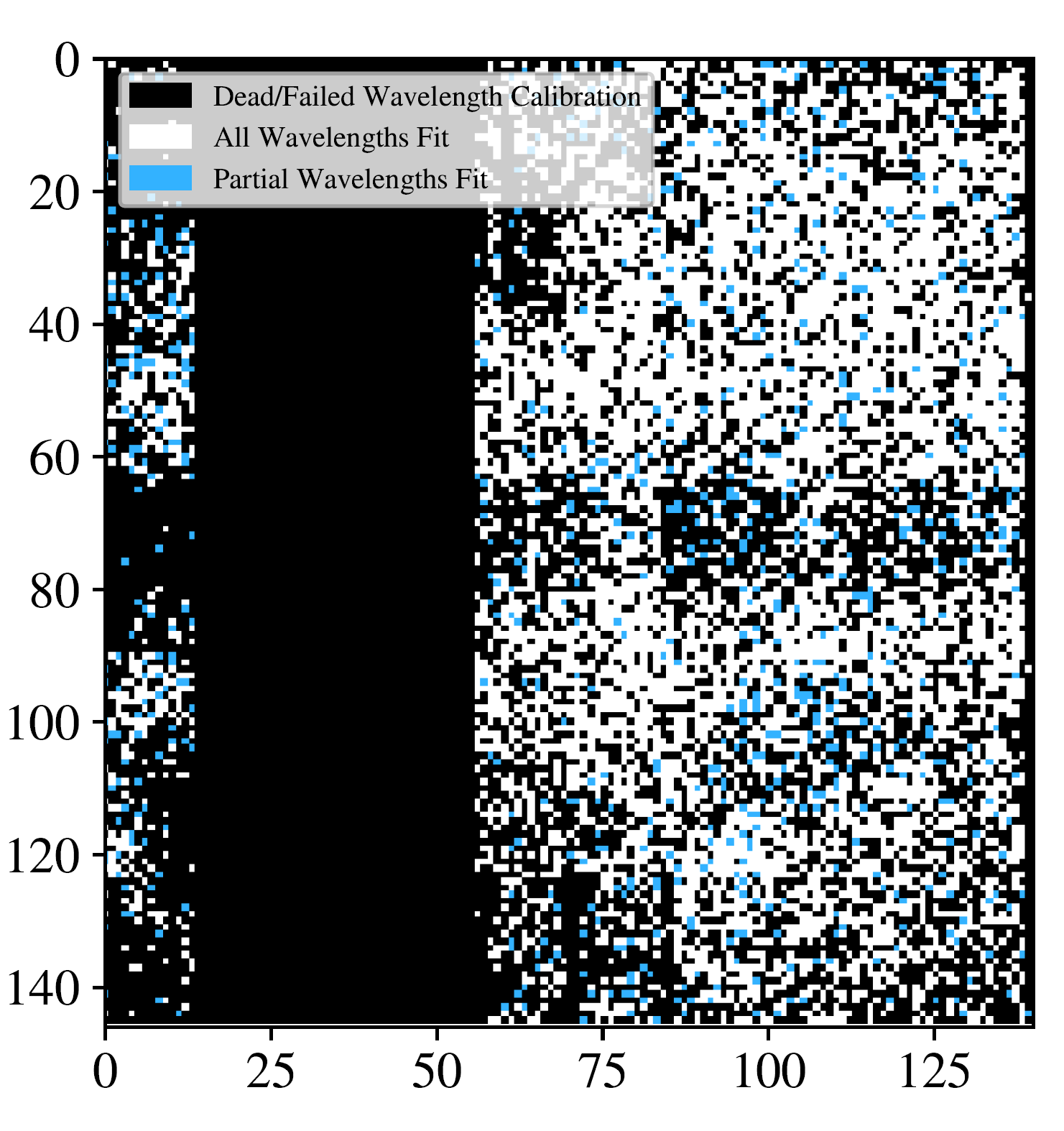}
  \caption{Wavelength calibration results for the engineering array in MEC. Fully wavelength calibrated pixels are shown in white, pixels calibrated using a subset of the available wavelengths are shown in blue, and pixels that had no wavelength calibration solution are shown in black.}
  \label{fig:wvlcal_map}
\end{figure}

The engineering MEC array pixel map is shown in Figure~\ref{fig:wvlcal_map}. For the fully working feedlines (6-10), 75\% of pixels are photosensitive with 61.4\% having a good wavelength calibration solution. When only looking at the best part of the detector (the ``sweet spot''), between 20 and 60 in the y direction, these numbers increase to 82.1\% and 74.3\%, respectively. The bad section in the center, between 60 and 80 in the y direction, is likely performing poorly as the device surface impedance was slightly higher than desired, so these resonators are lower frequency than designed (starting at 3.5 vs 3.9 GHz), causing some issues with resonator coupling to the feedline and HEMT gain.

\begin{figure}[t] 
  \centering
  \includegraphics[width=\columnwidth]{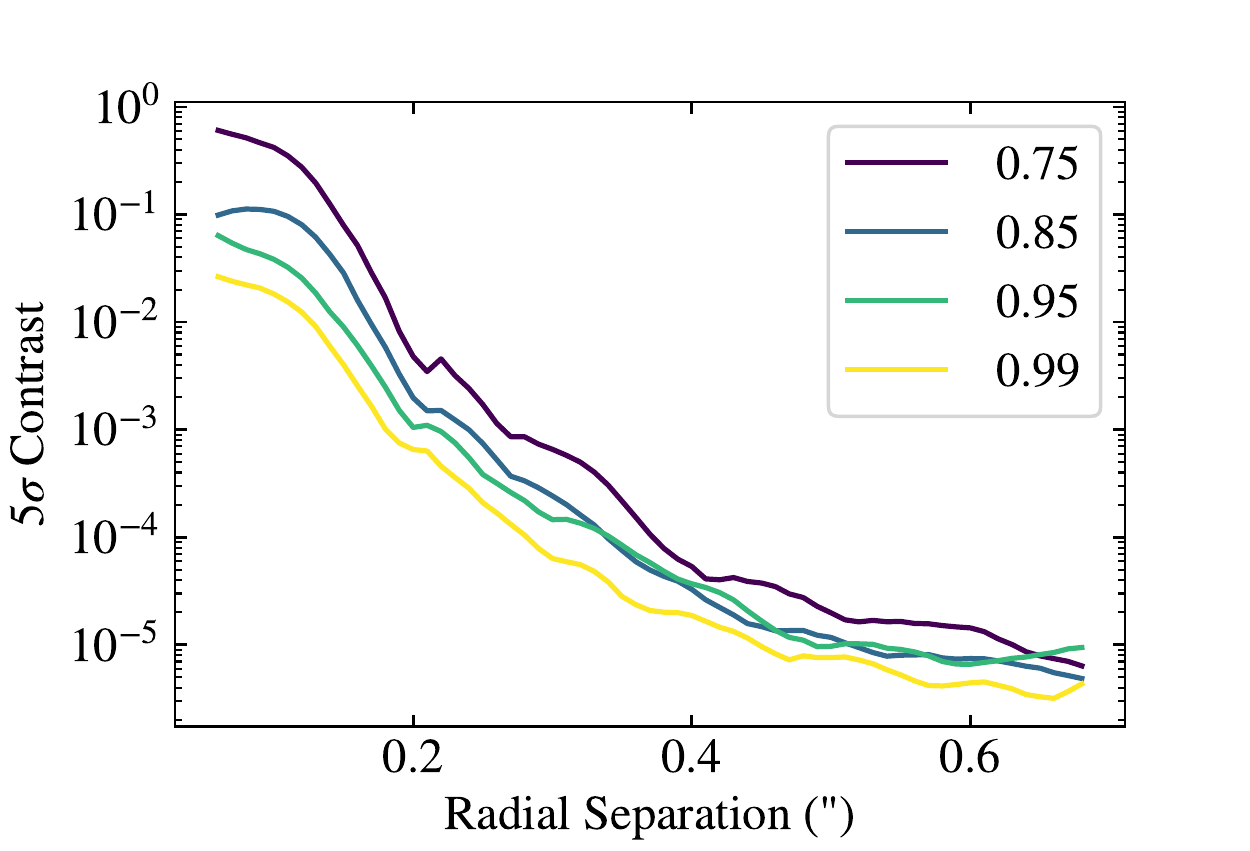}
  \caption{Simulation with the MEDIS package performed to evaluate the impact of dead pixels on the raw contrast of MEC.  The four different lines represent the effect of varying pixel yield on the $5 \sigma$ contrast.}
  \label{fig:MEDIS}
\end{figure}

Simulations with the MEDIS package~\citep{dodkins2020first} have been performed to estimate the impact of dead pixels on the raw contrast (before post processing with ADI/SDI/SSD/etc.) for exoplanet direct imaging. These simulations show that the raw contrast will be roughly two times worse with observed yield in the sweet spot compared with perfect yield, as shown in Figure~\ref{fig:MEDIS}. 

\subsection{Spectral Resolution}

\begin{figure}[!t] 
  \centering
  \includegraphics[width=\columnwidth]{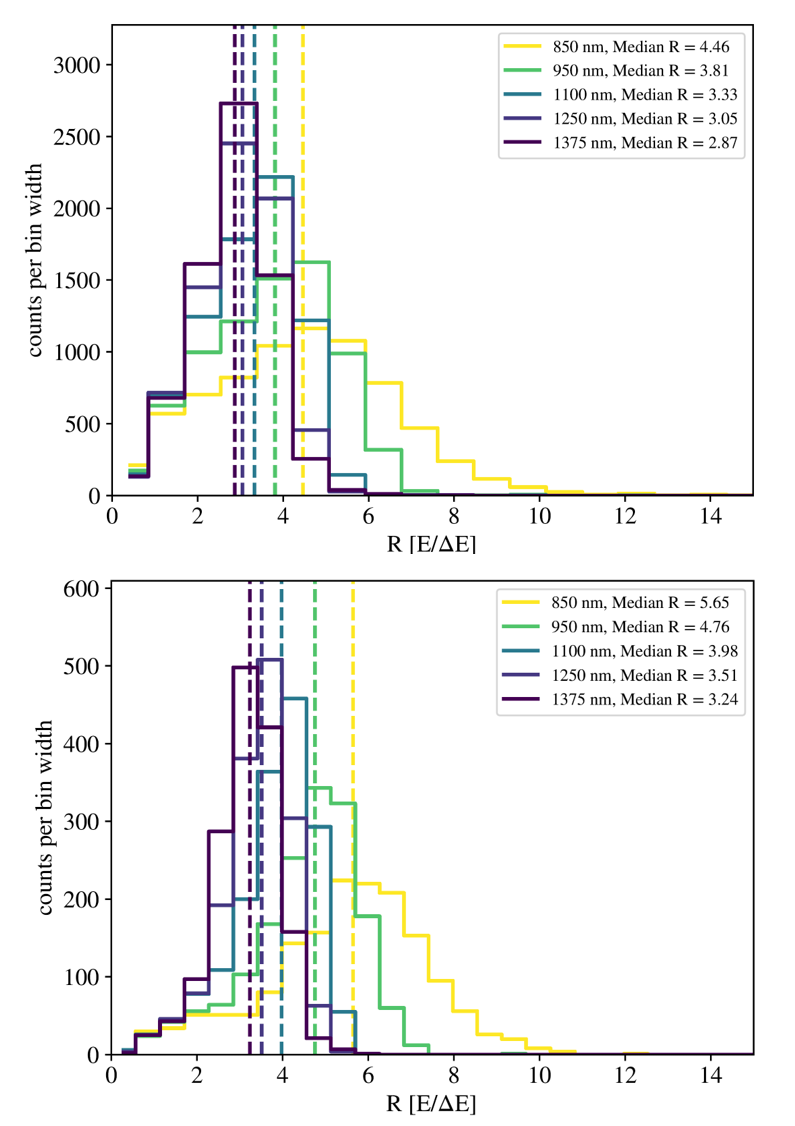}
  \caption{MEC spectral resolution for the whole array (top) and in the sweet spot (bottom). Note the differing vertical axes scales due to the smaller number of probed pixels in the best part of the array.}
  \label{fig:spectral_res_hist}
\end{figure}

\begin{figure}[!t] 
  \centering
  \includegraphics[width=\columnwidth]{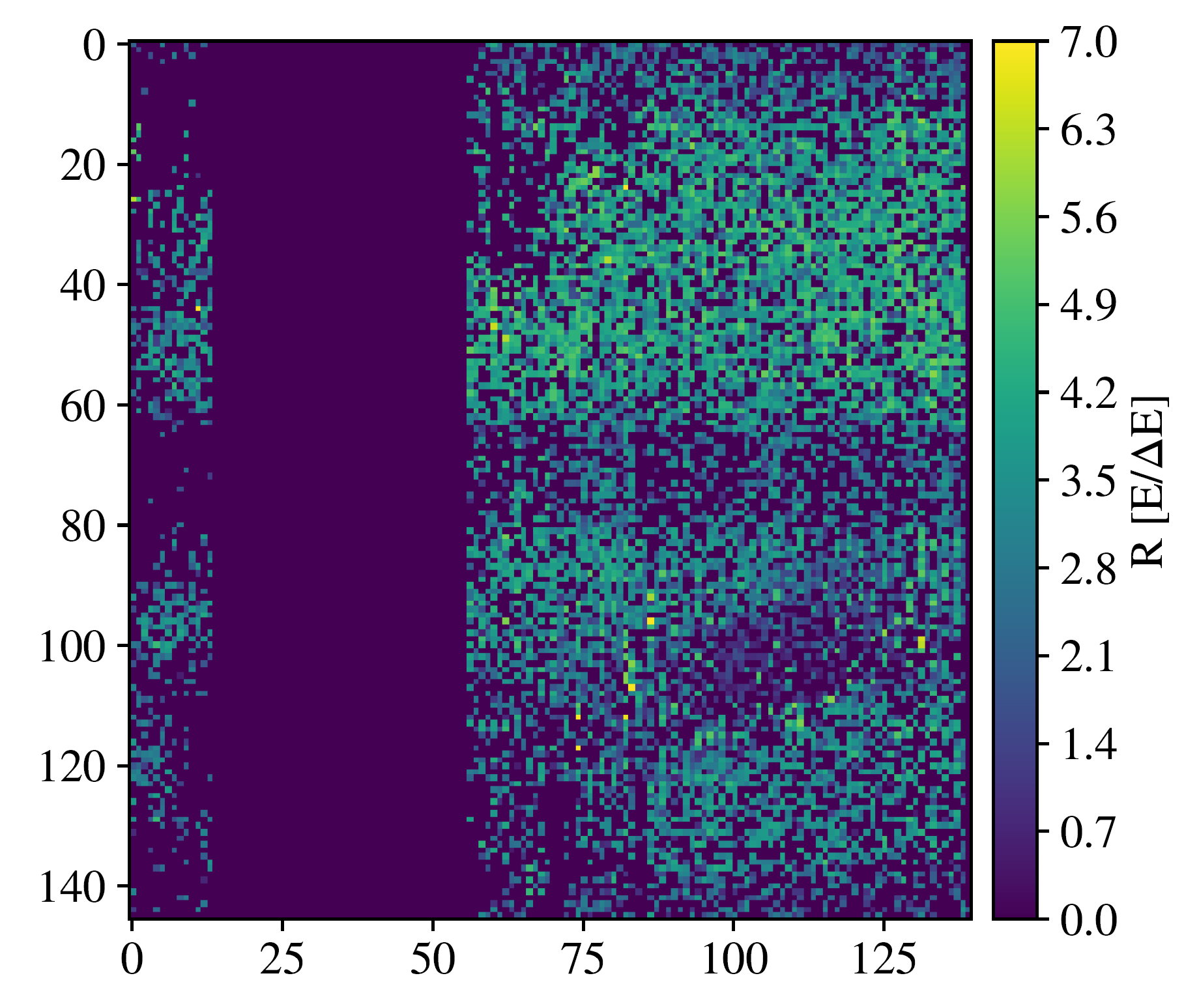}
  \caption{Map of energy resolution for MEC's engineering array at 1.1 $\mu m$}
  \label{fig:spectral_res_map}
\end{figure}

The spectral resolution of an MKID is a complex tradeoff between material, pixel geometry, and readout technique.  The spectral resolution of the MKIDs used in MEC are explored in great detail in \citet{2019ApPhL.115d2601Z}. However, this data was taken under ideal circumstances in a dilution refrigerator in our lab, not in an ADR at the top of a mountain on an electrically noisy Naysmth platform. Figure~\ref{fig:spectral_res_hist} shows the actual measured spectral resolution histograms of all the pixels in MEC, while Figure~\ref{fig:spectral_res_map} shows a map of spectral resolution at a single wavelength. 

A hand selected MKID pixel on a MEC-style PtSi array measured with a HEMT in the lab have a spectral resolution of around R=7.5 at 980 nm, while in MEC the median pixel in the good region of the array has R=5 at 980 nm. This degradation is likely related to several factors, including:

\begin{itemize}
    \item Excess noise in the readout caused primarily by loss of signal in the high density microwave cables~\citep{2020arXiv200706496S} that bring the signals to the HEMT.
    \item Intermodulation products causing unwanted tones that interfere with the pulse readout.
    \item Many pixels are not ideal due to fabrication errors, such as frequency overlaps.
\end{itemize}

The readout itself does not appear to introduce noticeable degradation aside from the intermodulation products, which can be reduced through an optimization algorithm~\citep{NeelayGen2}.

\subsection{Throughput}

Throughput was calculated by comparing a spectrum of the G0 type star BD +172803 attained by MEC with a calibrated spectrum of the same target. This calibrated spectrum was generated by taking a G0 type stellar spectrum from the PHEONIX library and normalizing it to match the reported J and H band flux for this object \citep{husser2013new, cutri20032mass}. In order to compare the two spectra, both were binned in 0.05 eV bins which over samples the median MKID energy resolution by roughly a factor of 4. 

\begin{figure}[!t] 
  \centering
  \includegraphics[width=\columnwidth]{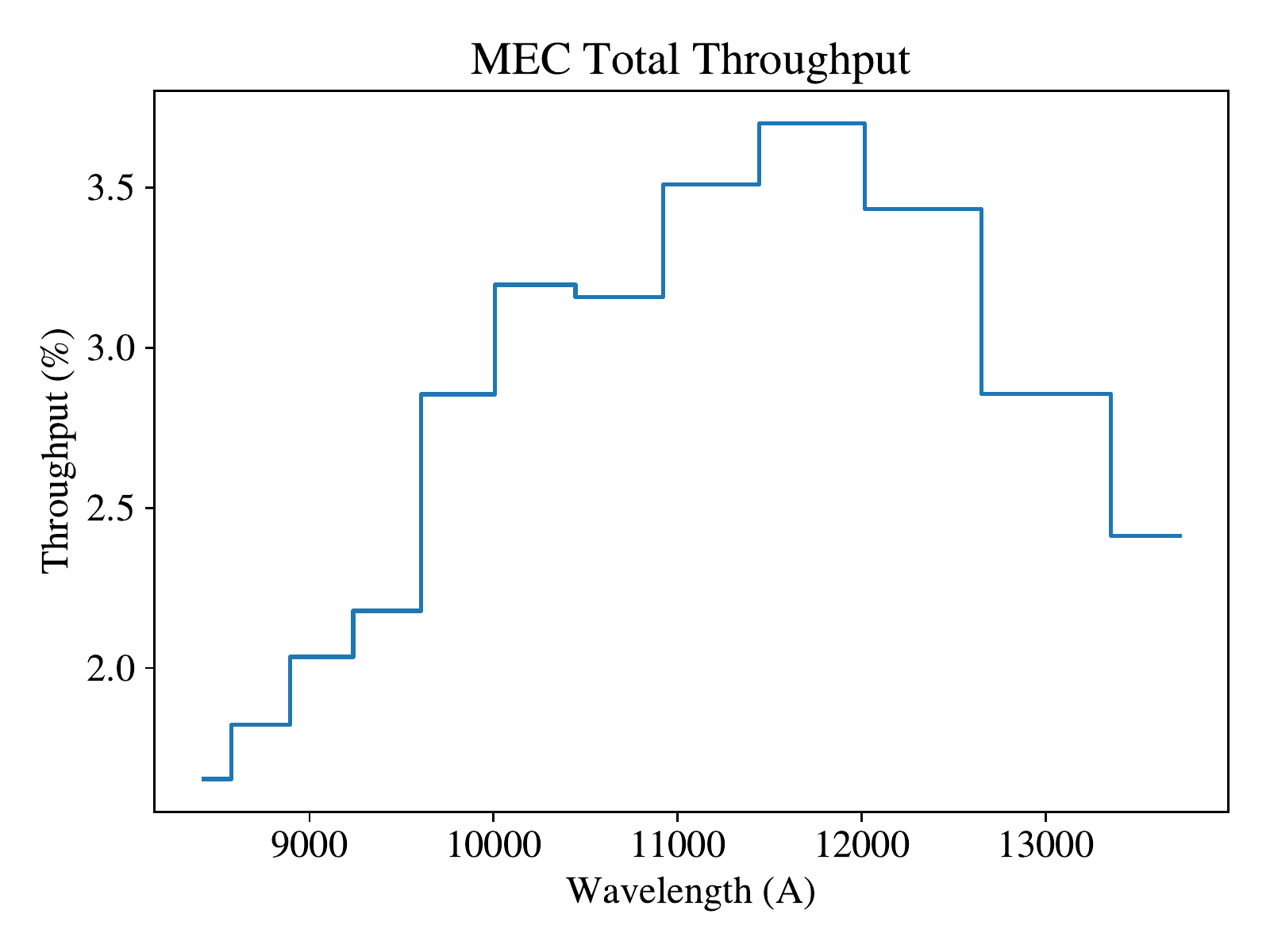}
  \caption{MEC throughput, from the top of the atmosphere to registered photons in the MKID detector, as a function of wavelength}
  \label{fig:spectral_resp_curve}
\end{figure}

Dividing this binned MEC spectrum by the binned calibrated spectrum should yield the total system throughput as a function of wavelength, but there are some additional factors that must first be taken into consideration. During this observation, a 90/10 beamsplitter upstream of the MEC optics in SCExAO was used which directed only 10\% of the total available light to MEC. Additionally, since this target has a J band magnitude of 6.6 and an H band magnitude of 6.4, MEC's 2.5 OD  filter had to be used so as to not saturate the array. This filter has not been measured as a function of wavelength for MEC and so represents a source of uncertainty in this calculation. Finally, the measured Strehl during this observation was found using the C-RED 2 IR camera in SCExAO to be 78\% in H band which must also be accounted for since a tight aperture was used so as to only encompass the core of the PSF. Since the Strehl is expected to be worse at shorter wavelengths, using this measured H band Strehl of 78\% represents a lower limit on the J band throughput.

Taking all of these factors into consideration, the resulting average J band throughput of the total optical system is 3.3$\pm$0.1\%, see Figure~\ref{fig:spectral_resp_curve}. This is roughly consistent with what we expect if the total throughput of the atmosphere, telescope, AO188, and SCExAO are roughly 20\%. It is possible slight misalignment of the microlens is reducing detector quantum efficiency, so careful measurements of the science grade array QE will be performed before installation to allow modelling of the full system thoughput to compare with on-sky measurements. 

\section{Future Plans}
\label{sec:fp}

\subsection{Upgrades to MEC}

Future hardware upgrades include a new anti-reflection coated science grade array which will roughly double the quantum efficiency and include fabrication process improvements that should greatly improve pixel yield. The MKID Pipeline will also be undergoing continual improvements which will allow us to more accurately determine the energy of each incident photon and to better remove false counts. For additional details on how future array improvements will affect the performance of MEC, see \citet{dodkins2020first}. 

\subsection{Wavefront Sensing and Control - Real-Time Integration with SCExAO}

MEC has already done on-sky speckle nulling using a code derived from ~\citet{2016SPIE.9909E..55B} and demonstrated significant suppression of the quasi-static speckle halo which will be published in a future paper.  This is the first step towards making MEC both a science camera and an effective focal plane wavefront sensor. Development is proceeding to speed speckle nulling up to allow suppression of not just quasi-static speckles, but fast atmospheric speckles, with expected full probing and cancellation cycles occurring at frequencies of at least 200 Hz. The implementation of more advanced coherent differential imaging (CDI) probing techniques~\citep{2017JATIS...3d5001M} are also being developed for use both in post-processing and in real time. 

In the long term, a fully optimized real time control package using the Frazin algorithm~\citep{2016JOSAA..33..712F,2018JOSAA..35..594F} or similar predictive control~\citep{2018JATIS...4a9001M} and sensor fusion approaches could help MEC use all available information simultaneously to approach fundamental photon noise limits and enable detection and characterization of smaller planets closer to their parent star. This technology will allow the next generation of instruments on 30-m class telescopes to potentially detect and characterize rocky planets in the habitable zones of nearby M-dwarfs.

\begin{figure}
\vspace{4pt}
  \includegraphics[width=\columnwidth]{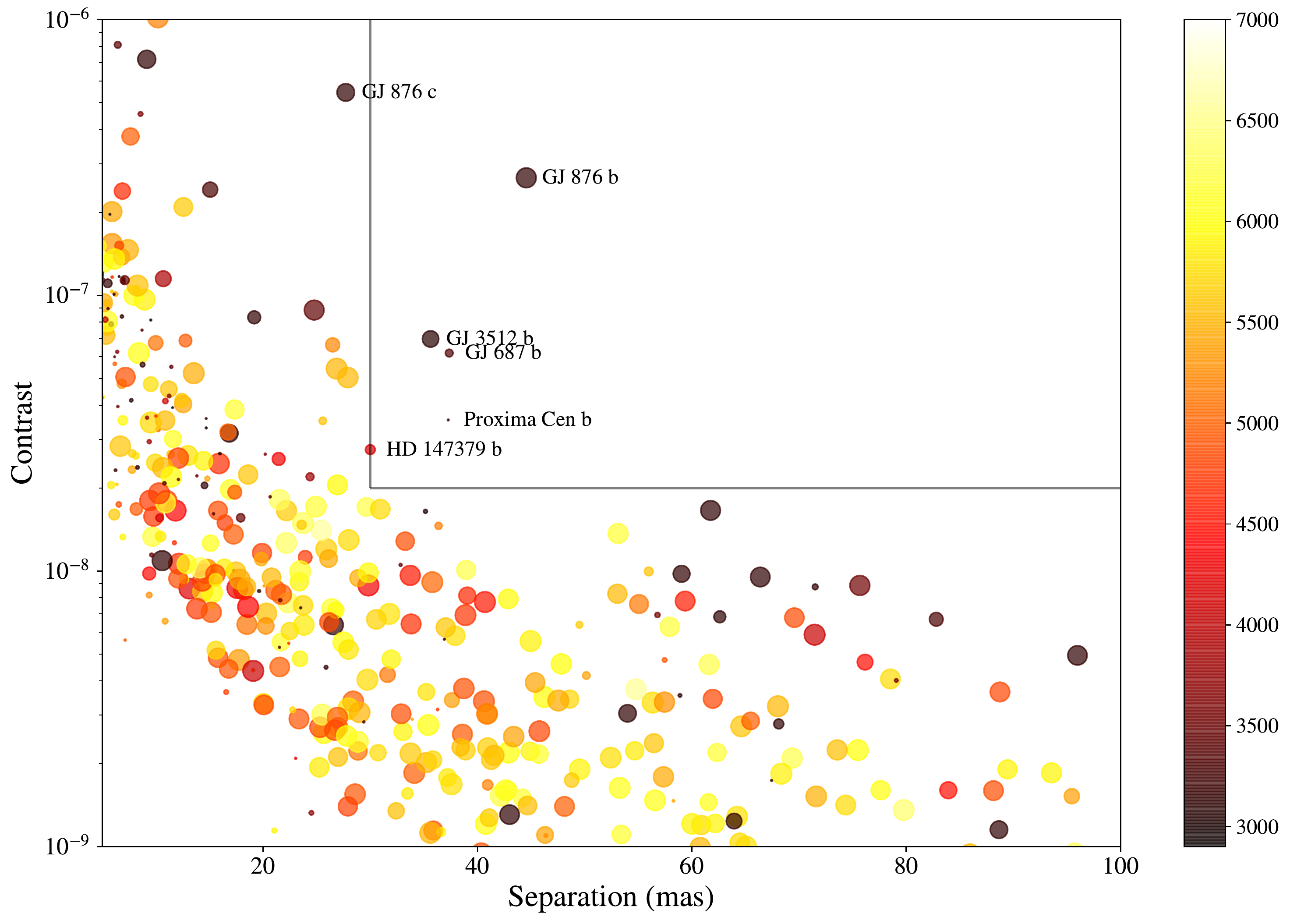}
  \caption{Angular separation vs. contrast for known exoplanets viewable with MEC/SCExAO and it's companion instrument DARKNESS~\citep{2018PASP..130f5001M} on MagAO-X. The size of the point represents the radius of the planet, and the color of the point represents the T$_{\mathrm{eff}}$ of the host star. The horizontal line at $2{\times}10^{-8}$ and the vertical line at 30 mas are meant to guide the eye to the most promising targets, which are labelled with their common names. Of these targets, GJ 876 b is by far the most promising.}
  \label{fig:refl}
\end{figure}

\subsection{Science Goals}

MEC was designed with the goal of enabling imaging of exoplanets in reflected light.  Figure~\ref{fig:refl} shows all currently known planets with their separation plotted against the expected contrast ratio. With MEC we hope to reach final contrasts of at $10^{-7}$ at 2 $\lambda/D$ after post-processing, enabling the imaging of at least GJ 876 b. This contrast limit is theoretically achievable by MEC given the inner working angle of the instrument and assuming we are photon noise dominated. Future papers, as mentioned above, will detail the use of both high speed realtime and post-processing techniques to reach this noise limit and ultimately enhance the discovery reach of SCExAO.

In addition to reflected light planets, MEC is also a powerful tool for the discovery and characterization of young giant planets still glowing from the heat of their formation. The shorter wavelengths MEC probes allows for superior angular resolution, and the high final contrast at small inner working angles, due to focal plane wavefront sensing and novel post-processing, should allow us to probe new parameter space.

\section{Acknowledgements}
Sarah Steiger and Neelay Fruitwala are supported by a grant from the Heising-Simons Foundation. Kristina K. Davis is supported by the NSF Astronomy and Astrophysics Postdoctoral Fellowship program under award number 1801983.  Jennifer P. Smith and Nicholas Zobrist are both supported by a NASA Space Technology Research Fellowship. Isabel Lipartito is supported by the National Science Foundation Graduate Research Fellowship under grant number 1650114. Frantz Martinache is funded by the European Research Council (ERC) under the European Union's Horizon 2020 research and innovation program (grant agreement CoG - 683029).

The development of SCExAO was supported by the Japan Society for the Promotion of Science (Grant-in-Aid for Research \#23340051, \#26220704, \#23103002, \#19H00703 \& \#19H00695), the Astrobiology Center of the National Institutes of Natural Sciences, Japan, the Mt. Cuba Foundation and the director’s contingency fund at Subaru Telescope. The authors wish to recognize and acknowledge the very significant cultural role and reverence that the summit of Maunakea has always had within the indigenous Hawaiian community, and are most fortunate to have the opportunity to conduct observations from this mountain.

\bibliography{bib1.bib}

\begin{thebibliography}{}
\expandafter\ifx\csname natexlab\endcsname\relax\def\natexlab#1{#1}\fi
\providecommand{\url}[1]{\href{#1}{#1}}
\providecommand{\dodoi}[1]{doi:~\href{http://doi.org/#1}{\nolinkurl{#1}}}
\providecommand{\doeprint}[1]{\href{http://ascl.net/#1}{\nolinkurl{http://ascl.net/#1}}}
\providecommand{\doarXiv}[1]{\href{https://arxiv.org/abs/#1}{\nolinkurl{https://arxiv.org/abs/#1}}}

\bibitem[{Beuzit {et~al.}(2019)Beuzit, Vigan, Mouillet, Dohlen, Gratton,
  Boccaletti, Sauvage, Schmid, Langlois, Petit, {et~al.}}]{beuzit2019sphere}
Beuzit, J.-L., Vigan, A., Mouillet, D., {et~al.} 2019, Astronomy \&
  Astrophysics, 631, A155

\bibitem[{Bottom {et~al.}(2016)Bottom, Femenia, Huby, Mawet, Dekany, Milburn,
  \& Serabyn}]{2016SPIE.9909E..55B}
Bottom, M., Femenia, B., Huby, E., {et~al.} 2016, in Proceedings of the SPIE,
  California Institute of Technology (United States) (International Society for
  Optics and Photonics), 990955

\bibitem[{Bowler(2016)}]{Bowler2016}
Bowler, B.~P. 2016, PASP, 128, 102001

\bibitem[{Carbillet {et~al.}(2011)Carbillet, Bendjoya, Abe, Guerri, Boccaletti,
  Daban, Dohlen, Ferrari, Robbe-Dubois, Douet, \& Vakili}]{Carbillet2011}
Carbillet, M., Bendjoya, P., Abe, L., {et~al.} 2011, Experimental Astronomy,
  30, 39, \dodoi{10.1007/s10686-011-9219-4}

\bibitem[{Close {et~al.}(2013)Close, Males, Morzinski, Kopon, Follette,
  Rodigas, Hinz, Wu, Puglisi, Esposito, Riccardi, Pinna, Xompero, Briguglio,
  Uomoto, \& Hare}]{Close_2013}
Close, L.~M., Males, J.~R., Morzinski, K., {et~al.} 2013, ApJ, 774, 94

\bibitem[{Crepp {et~al.}(2011)Crepp, Pueyo, Brenner, Oppenheimer, Zimmerman,
  Hinkley, Parry, King, Vasisht, Beichman, Hillenbrand, Dekany, Shao, Burruss,
  Roberts, Bouchez, Roberts, \& Soummer}]{Crepp2011}
Crepp, J.~R., Pueyo, L., Brenner, D., {et~al.} 2011, ApJ, 729, 132

\bibitem[{Currie {et~al.}(2014)Currie, Daemgen, Debes, Lafreniere, Itoh,
  Jayawardhana, Ratzka, \& Correia}]{currie2013direct}
Currie, T., Daemgen, S., Debes, J., {et~al.} 2014, The Astrophysical Journal
  Letters, 780, L30

\bibitem[{Cutri {et~al.}(2003)Cutri, Skrutskie, Van~Dyk, Beichman, Carpenter,
  Chester, Cambresy, Evans, Fowler, Gizis, {et~al.}}]{cutri20032mass}
Cutri, R., Skrutskie, M., Van~Dyk, S., {et~al.} 2003, tmc

\bibitem[{Day {et~al.}(2003)Day, Leduc, A~Mazin, Vayonakis, \&
  Zmuidzinas}]{Day_2003}
Day, P., Leduc, H., A~Mazin, B., Vayonakis, A., \& Zmuidzinas, J. 2003, Nature,
  425, 817, \dodoi{10.1038/nature02037}

\bibitem[{Dodkins {et~al.}(2018)Dodkins, Mahashabde, O'Brien, Thatte,
  Fruitwala, Walter, Meeker, Szypryt, \& Mazin}]{2018A&C....23...60D}
Dodkins, R., Mahashabde, S., O'Brien, K., {et~al.} 2018, Astronomy and
  Computing, 23, 60

\bibitem[{Dodkins {et~al.}(2020)Dodkins, Davis, Lewis, Mahashabde, Mazin,
  Lipartito, Fruitwala, O’Brien, \& Thatte}]{dodkins2020first}
Dodkins, R.~H., Davis, K.~K., Lewis, B., {et~al.} 2020, Publications of the
  Astronomical Society of the Pacific, 132, 104503

\bibitem[{Frazin(2016)}]{2016JOSAA..33..712F}
Frazin, R.~A. 2016, Journal of the Optical Society of America A, 33, 712

\bibitem[{Frazin(2018)}]{2018JOSAA..35..594F}
---. 2018, Journal of the Optical Society of America A, 35, 594

\bibitem[{Fruitwala(Submitted)}]{NeelayGen2}
Fruitwala, N. Submitted, AIP Review of Scientific Instruments

\bibitem[{Gerard {et~al.}(2019)Gerard, Marois, Currie, Brandt, Chilcote,
  Draper, Groff, Guyon, Hayashi, Jovanovic, Knapp, Kudo, Kwon, Lozi,
  Martinache, McElwain, Tamura, \& Uyama}]{Gerard+Marois+Currie+etal_2019}
Gerard, B.~L., Marois, C., Currie, T., {et~al.} 2019, ApJ, 158, 36

\bibitem[{{Gomez Gonzalez, C. A.} {et~al.}(2016){Gomez Gonzalez, C. A.},
  {Absil, O.}, {Absil, P.-A.}, {Van Droogenbroeck, M.}, {Mawet, D.}, \&
  {Surdej, J.}}]{Gonzalez_2016}
{Gomez Gonzalez, C. A.}, {Absil, O.}, {Absil, P.-A.}, {et~al.} 2016, A\&A, 589,
  A54

\bibitem[{Guyon(2005)}]{Guyon_2005}
Guyon, O. 2005, ApJ, 629, 592

\bibitem[{Hodapp {et~al.}(2008)Hodapp, Suzuki, Tamura, Abe, Suto, Kandori,
  Morino, Nishimura, Takami, Guyon, {et~al.}}]{hodapp2008hiciao}
Hodapp, K.~W., Suzuki, R., Tamura, M., {et~al.} 2008, in Ground-based and
  Airborne Instrumentation for Astronomy II, Vol. 7014, International Society
  for Optics and Photonics, 701419

\bibitem[{Husser {et~al.}(2013)Husser, Wende-von Berg, Dreizler, Homeier,
  Reiners, Barman, \& Hauschildt}]{husser2013new}
Husser, T.-O., Wende-von Berg, S., Dreizler, S., {et~al.} 2013, Astronomy \&
  Astrophysics, 553, A6

\bibitem[{Jovanovic {et~al.}(2015)Jovanovic, Martinache, Guyon, Clergeon,
  Singh, Kudo, Garrel, Newman, Doughty, Lozi, Males, Minowa, Hayano, Takato,
  Morino, Kuhn, Serabyn, Norris, Tuthill, Schworer, Stewart, Close, Huby,
  Perrin, Lacour, Gauchet, Vievard, Murakami, Oshiyama, Baba, Matsuo,
  Nishikawa, Tamura, Lai, Marchis, Duchene, Kotani, \& Woillez}]{Jovanovic2015}
Jovanovic, N., Martinache, F., Guyon, O., {et~al.} 2015, PASP, 127, 890

\bibitem[{Jovanovic {et~al.}(2019)Jovanovic, Delorme, Bond, Cetre, Mawet,
  Echeverri, Wallace, Bartos, Lilley, Ragland, {et~al.}}]{jovanovic2019keck}
Jovanovic, N., Delorme, J.-R., Bond, C.~Z., {et~al.} 2019, arXiv preprint
  arXiv:1909.04541

\bibitem[{{Keppler} {et~al.}(2018){Keppler}, {Benisty}, {M{\"u}ller},
  {Henning}, {van Boekel}, {Cantalloube}, {Ginski}, {van Holstein}, {Maire},
  {Pohl}, {Samland }, {Avenhaus}, {Baudino}, {Boccaletti}, {de Boer},
  {Bonnefoy}, {Chauvin}, {Desidera}, {Langlois}, {Lazzoni}, {Marleau},
  {Mordasini}, {Pawellek}, {Stolker}, {Vigan}, {Zurlo}, {Birnstiel},
  {Brandner}, {Feldt}, {Flock}, {Girard}, {Gratton}, {Hagelberg}, {Isella},
  {Janson}, {Juhasz}, {Kemmer}, {Kral}, {Lagrange}, {Launhardt}, {Matter},
  {M{\'e}nard}, {Milli}, {Molli{\`e}re}, {Olofsson}, {P{\'e}rez}, {Pinilla},
  {Pinte}, {Quanz}, {Schmidt}, {Udry}, {Wahhaj}, {Williams}, {Buenzli},
  {Cudel}, {Dominik}, {Galicher}, {Kasper}, {Lannier}, {Mesa}, {Mouillet},
  {Peretti}, {Perrot}, {Salter}, {Sissa}, {Wildi}, {Abe}, {Antichi},
  {Augereau}, {Baruffolo}, {Baudoz}, {Bazzon}, {Beuzit}, {Blanchard}, {Brems},
  {Buey}, {De Caprio}, {Carbillet}, {Carle}, {Cascone}, {Cheetham}, {Claudi},
  {Costille}, {Delboulb{\'e}}, {Dohlen}, {Fantinel}, {Feautrier}, {Fusco},
  {Giro}, {Gluck}, {Gry}, {Hubin}, {Hugot}, {Jaquet}, {Le Mignant}, {Llored},
  {Madec}, {Magnard}, {Martinez}, {Maurel}, {Meyer}, {M{\"o}ller-Nilsson},
  {Moulin}, {Mugnier}, {Orign{\'e}}, {Pavlov}, {Perret}, {Petit}, {Pragt},
  {Puget}, {Rabou}, {Ramos}, {Rigal}, {Rochat}, {Roelfsema}, {Rousset}, {Roux},
  {Salasnich}, {Sauvage}, {Sevin}, {Soenke}, {Stadler}, {Suarez}, {Turatto}, \&
  {Weber}}]{Keppler+Benisty+Muller+etal_2018}
{Keppler}, M., {Benisty}, M., {M{\"u}ller}, A., {et~al.} 2018, A\&A, 617, A44

\bibitem[{{Kuzuhara} {et~al.}(2013){Kuzuhara}, {Tamura}, {Kudo}, {Janson},
  {Kand ori}, {Brandt}, {Thalmann}, {Spiegel}, {Biller}, {Carson}, {Hori},
  {Suzuki}, {Burrows}, {Henning}, {Turner}, {McElwain}, {Moro-Mart{\'\i}n},
  {Suenaga}, {Takahashi}, {Kwon}, {Lucas}, {Abe}, {Brand ner}, {Egner},
  {Feldt}, {Fujiwara}, {Goto}, {Grady}, {Guyon}, {Hashimoto}, {Hayano},
  {Hayashi}, {Hayashi}, {Hodapp}, {Ishii}, {Iye}, {Knapp}, {Matsuo}, {Mayama},
  {Miyama}, {Morino}, {Nishikawa}, {Nishimura}, {Kotani}, {Kusakabe}, {Pyo},
  {Serabyn}, {Suto}, {Takami}, {Takato}, {Terada}, {Tomono}, {Watanabe},
  {Wisniewski}, {Yamada}, {Takami}, \&
  {Usuda}}]{Kuzuhara+Tamura+Kudo+etal_2013}
{Kuzuhara}, M., {Tamura}, M., {Kudo}, T., {et~al.} 2013, ApJ, 774, 11

\bibitem[{{Lafreni{\`e}re} {et~al.}(2007){Lafreni{\`e}re}, {Marois}, {Doyon},
  {Nadeau}, \& {Artigau}}]{Lafreniere+Marois+Doyon+etal_2007}
{Lafreni{\`e}re}, D., {Marois}, C., {Doyon}, R., {Nadeau}, D., \& {Artigau},
  {\'E}. 2007, ApJ, 660, 770

\bibitem[{{Lagrange} {et~al.}(2010){Lagrange}, {Bonnefoy}, {Chauvin}, {Apai},
  {Ehrenreich}, {Boccaletti}, {Gratadour}, {Rouan}, {Mouillet}, {Lacour}, \&
  {Kasper}}]{Lagrange+Bonnefoy+Chauvin+etal_2010}
{Lagrange}, A.~M., {Bonnefoy}, M., {Chauvin}, G., {et~al.} 2010, Science, 329,
  57

\bibitem[{Lewis \& Oppenheimer(2017)}]{lewis2017direct}
Lewis, B., \& Oppenheimer, R. 2017, Columbia Undergraduate Science Journal, 11

\bibitem[{Lozi {et~al.}(2018)Lozi, Guyon, Jovanovic, Takato, Singh, Norris,
  Okita, Bando, \& Martinache}]{Lozi_2018}
Lozi, J., Guyon, O., Jovanovic, N., {et~al.} 2018, Journal of Astronomical
  Telescopes, Instruments, and Systems, 4, 1

\bibitem[{Macintosh {et~al.}(2014)Macintosh, Graham, Ingraham, Konopacky,
  Marois, Perrin, Poyneer, Bauman, Barman, Burrows, Cardwell, Chilcote,
  De~Rosa, Dillon, Doyon, Dunn, Erikson, Fitzgerald, Gavel, Goodsell, Hartung,
  Hibon, Kalas, Larkin, Maire, Marchis, Marley, McBride, Millar-Blanchaer,
  Morzinski, Norton, Oppenheimer, Palmer, Patience, Pueyo, Rantakyro, Sadakuni,
  Saddlemyer, Savransky, Serio, Soummer, Sivaramakrishnan, Song, Thomas,
  Wallace, Wiktorowicz, \& Wolff}]{Macintosh2014}
Macintosh, B., Graham, J.~R., Ingraham, P., {et~al.} 2014, Proceedings of the
  National Academy of Sciences, 111, 12661, \dodoi{10.1073/pnas.1304215111}

\bibitem[{{Macintosh} {et~al.}(2015){Macintosh}, {Graham}, {Barman}, {De Rosa},
  {Konopacky}, {Marley}, {Marois}, {Nielsen}, {Pueyo}, {Rajan}, {Rameau},
  {Saumon}, {Wang}, {Patience}, {Ammons}, {Arriaga}, {Artigau}, {Beckwith},
  {Brewster}, {Bruzzone}, {Bulger}, {Burningham}, {Burrows}, {Chen}, {Chiang},
  {Chilcote}, {Dawson}, {Dong}, {Doyon}, {Draper}, {Duch{\^e}ne}, {Esposito},
  {Fabrycky}, {Fitzgerald}, {Follette}, {Fortney}, {Gerard}, {Goodsell},
  {Greenbaum}, {Hibon}, {Hinkley}, {Cotten}, {Hung}, {Ingraham},
  {Johnson-Groh}, {Kalas}, {Lafreniere}, {Larkin}, {Lee}, {Line}, {Long},
  {Maire}, {Marchis}, {Matthews}, {Max}, {Metchev}, {Millar-Blanchaer},
  {Mittal}, {Morley}, {Morzinski}, {Murray-Clay}, {Oppenheimer}, {Palmer},
  {Patel}, {Perrin}, {Poyneer}, {Rafikov}, {Rantakyr{\"o}}, {Rice}, {Rojo},
  {Rudy}, {Ruffio}, {Ruiz}, {Sadakuni}, {Saddlemyer}, {Salama}, {Savransky},
  {Schneider}, {Sivaramakrishnan}, {Song}, {Soummer}, {Thomas}, {Vasisht},
  {Wallace}, {Ward-Duong}, {Wiktorowicz}, {Wolff}, \&
  {Zuckerman}}]{Macintosh+Graham+Barman+etal_2015}
{Macintosh}, B., {Graham}, J.~R., {Barman}, T., {et~al.} 2015, Science, 350, 64

\bibitem[{Macintosh {et~al.}(2008)Macintosh, Grahamand, Palmer, Doyon, Dunn,
  Gavel, Larkin, Oppenheimer, Saddlemyer, Sivaramakrishnan, Wallace, Bauman,
  Erickson, Marois, Poyneer, \& Soummer}]{Macintosh2008}
Macintosh, B.~A., Grahamand, J.~R., Palmer, D.~W., {et~al.} 2008, SPIE, 7015,
  7015

\bibitem[{Males \& Guyon(2018)}]{2018JATIS...4a9001M}
Males, J.~R., \& Guyon, O. 2018, Journal of Astronomical Telescopes, 4, 019001

\bibitem[{Males {et~al.}(2018)Males, Close, Miller, Schatz, Doelman, Lumbres,
  Snik, Rodack, Knight, Gorkom, Long, Hedglen, Kautz, Jovanovic, Morzinski,
  Guyon, Douglas, Follette, Lozi, Bohlman, Durney, Gasho, Hinz, Ireland, Jean,
  Keller, Kenworthy, Mazin, Noenickx, Alfred, Perez, Sanchez, Sauve,
  Weinberger, \& Conrad}]{Males2018}
Males, J.~R., Close, L.~M., Miller, K., {et~al.} 2018, SPIE, 10703

\bibitem[{{Marois} {et~al.}(2014){Marois}, {Correia}, {Galicher}, {Ingraham},
  {Macintosh}, {Currie}, \& {De Rosa}}]{Marois+Correia+Galicher+etal_2014}
{Marois}, C., {Correia}, C., {Galicher}, R., {et~al.} 2014, in SPIE, Vol. 9148,
  Adaptive Optics Systems IV, 91480U

\bibitem[{{Marois} {et~al.}(2000){Marois}, {Doyon}, {Racine}, \&
  {Nadeau}}]{Marois+Doyon+Racine+etal_2000}
{Marois}, C., {Doyon}, R., {Racine}, R., \& {Nadeau}, D. 2000, PASP, 112, 91

\bibitem[{{Marois} {et~al.}(2006){Marois}, {Lafreni{\`e}re}, {Doyon},
  {Macintosh}, \& {Nadeau}}]{Marois+Lafreniere+Doyon+etal_2006}
{Marois}, C., {Lafreni{\`e}re}, D., {Doyon}, R., {Macintosh}, B., \& {Nadeau},
  D. 2006, ApJ, 641, 556

\bibitem[{{Marois} {et~al.}(2008){Marois}, {Macintosh}, {Barman}, {Zuckerman},
  {Song}, {Patience}, {Lafreni{\`e}re}, \&
  {Doyon}}]{Marois+Macintosh+Barman+etal_2008}
{Marois}, C., {Macintosh}, B., {Barman}, T., {et~al.} 2008, Science, 322, 1348

\bibitem[{Matthews {et~al.}(2017)Matthews, Crepp, Vasisht, \&
  Cady}]{2017JATIS...3d5001M}
Matthews, C.~T., Crepp, J.~R., Vasisht, G., \& Cady, E. 2017, Journal of
  Astronomical Telescopes, 3, 045001

\bibitem[{Mawet {et~al.}(2014{\natexlab{a}})Mawet, Shelton, Wallace, Bottom,
  Kuhn, Mennesson, Burruss, Bartos, Pueyo, Carlotti, \& Serabyn}]{Mawet2014}
Mawet, D., Shelton, C., Wallace, J., {et~al.} 2014{\natexlab{a}}, SPIE, 9143,
  9143

\bibitem[{Mawet {et~al.}(2014{\natexlab{b}})Mawet, Milli, Wahhaj, Pelat, Absil,
  Delacroix, Boccaletti, Kasper, Kenworthy, Marois, Mennesson, \&
  Pueyo}]{Mawet_2014}
Mawet, D., Milli, J., Wahhaj, Z., {et~al.} 2014{\natexlab{b}}, ApJ, 792, 97

\bibitem[{Mawet {et~al.}(2016)Mawet, Wizinowich, Dekany, Chun, Hall, Cetre,
  Guyon, Wallace, Bowler, Liu, Ruane, Serabyn, Bartos, Wang, Vasisht,
  Fitzgerald, Skemer, Ireland, Fucik, Fortney, Crossfield, Hu, \&
  Benneke}]{Mawet2016}
Mawet, D., Wizinowich, P., Dekany, R., {et~al.} 2016, SPIE, 9909, 9909

\bibitem[{Mazin {et~al.}(2012)Mazin, Bumble, Meeker, O'Brien, McHugh, \&
  Langman}]{Mazin_2012}
Mazin, B.~A., Bumble, B., Meeker, S.~R., {et~al.} 2012, Opt. Express, 20, 1503

\bibitem[{Mazin {et~al.}(2013)Mazin, Meeker, Strader, Szypryt, Marsden, van
  Eyken, Duggan, Walter, Ulbricht, Johnson, Bumble, O’Brien, \&
  Stoughton}]{ARCONS}
Mazin, B.~A., Meeker, S.~R., Strader, M.~J., {et~al.} 2013, PASP, 125, 1348

\bibitem[{McHugh {et~al.}(2012)McHugh, Mazin, Serfass, Meeker, O’Brien, Duan,
  Raffanti, \& Werthimer}]{McHugh_2012}
McHugh, S., Mazin, B.~A., Serfass, B., {et~al.} 2012, Review of Scientific
  Instruments, 83, 044702, \dodoi{10.1063/1.3700812}

\bibitem[{Meeker {et~al.}(2018{\natexlab{a}})Meeker, Mazin, Walter, Strader,
  Fruitwala, Bockstiegel, Szypryt, Ulbricht, Coiffard, Bumble, Cancelo, Zmuda,
  Treptow, Wilcer, Collura, Dodkins, Lipartito, Zobrist, Bottom, Shelton,
  Mawet, {Van Eyken}, Vasisht, \& Serabyn}]{Meeker2018}
Meeker, S.~R., Mazin, B.~A., Walter, A.~B., {et~al.} 2018{\natexlab{a}}, PASP,
  130

\bibitem[{Meeker {et~al.}(2018{\natexlab{b}})Meeker, Mazin, Walter, Strader,
  Fruitwala, Bockstiegel, Szypryt, Ulbricht, Coiffard, Bumble, Cancelo, Zmuda,
  Treptow, Wilcer, Collura, Dodkins, Lipartito, Zobrist, Bottom, Shelton,
  Mawet, van Eyken, Vasisht, \& Serabyn}]{2018PASP..130f5001M}
---. 2018{\natexlab{b}}, Publications of the Astronomical Society of the
  Pacific, 130, 065001

\bibitem[{{Racine} {et~al.}(1999){Racine}, {Walker}, {Nadeau}, {Doyon}, \&
  {Marois}}]{Racine+Walker+Nadeau+etal_1999}
{Racine}, R., {Walker}, G. A.~H., {Nadeau}, D., {Doyon}, R., \& {Marois}, C.
  1999, PASP, 111, 587

\bibitem[{{Smith} {et~al.}(2020){Smith}, {Mazin}, {Walter}, {Daal}, {Bailey
  III}, {Bockstiegel}, {Zobrist}, {Swimmer}, {Steiger}, \&
  {Fruitwala}}]{2020arXiv200706496S}
{Smith}, J.~P., {Mazin}, B.~A., {Walter}, A.~B., {et~al.} 2020, IEEE
  Transactions on Applied Superconductivity, 31, 1

\bibitem[{{Soummer} {et~al.}(2012){Soummer}, {Pueyo}, \&
  {Larkin}}]{Soummer+Pueyo+Larkin_2012}
{Soummer}, R., {Pueyo}, L., \& {Larkin}, J. 2012, ApJ, 755, L28

\bibitem[{{Sparks} \& {Ford}(2002)}]{Sparks+Ford_2002}
{Sparks}, W.~B., \& {Ford}, H.~C. 2002, ApJ, 578, 543

\bibitem[{Strader(2016)}]{Strader_2016}
Strader, M. 2016, PhD thesis, University of California Santa Barbara

\bibitem[{Szypryt {et~al.}(2017)Szypryt, Meeker, Coiffard, Fruitwala, Bumble,
  Ulbricht, Walter, Daal, Bockstiegel, Collura, Zobrist, Lipartito, \&
  Mazin}]{Szypryt_2017}
Szypryt, P., Meeker, S.~R., Coiffard, G., {et~al.} 2017, Opt. Express, 25,
  25894

\bibitem[{van Eyken {et~al.}(2015)van Eyken, Strader, Walter, Meeker, Szypryt,
  Stoughton, O'Brien, Marsden, Rice, Lin, \& Mazin}]{Eyken_2015}
van Eyken, J.~C., Strader, M.~J., Walter, A.~B., {et~al.} 2015, ApJS, 219, 14

\bibitem[{Walter {et~al.}(2019)Walter, Bockstiegel, Brandt, \&
  Mazin}]{Walter2019a}
Walter, A.~B., Bockstiegel, C., Brandt, T.~D., \& Mazin, B.~A. 2019,
  Publications of the Astronomy Society of the Pacific, 131, 114506

\bibitem[{Zobrist {et~al.}(2019)Zobrist, Eom, Day, Mazin, Meeker, Bumble,
  Leduc, Coiffard, Szypryt, Fruitwala, Lipartito, \&
  Bockstiegel}]{2019ApPhL.115d2601Z}
Zobrist, N., Eom, B.~H., Day, P., {et~al.} 2019, Applied Physics Letters, 115,
  042601

\end{thebibliography}
\bibliographystyle{aasjournal}

\end{document}